\documentclass[pra,aps,groupaddress]{revtex4}
\usepackage{amsfonts}
\usepackage{epsfig,amsmath,graphicx,amssymb,overpic,bm}
\usepackage{color,tikz,xcolor,extarrows}

\def\be{\begin{equation}}
\def\ee{\end{equation}}
\def\bee{\begin{eqnarray}}
\def\ene{\end{eqnarray}}
\def\bes{\begin{subequations}}
\def\ees{\end{subequations}}
\def\no{\nonumber}
\def\Re{{\rm Re}\,}
\def\Im{{\rm Im}\,}
\def\d{\displaystyle}

\begin{document}

\title{An initial-boundary value problem for the integrable spin-1 Gross-Pitaevskii equations with a $4\times 4$ Lax pair on the half-line}
\author{Zhenya Yan}
\email{zyyan@mmrc.iss.ac.cn}
\affiliation{\vspace{0.1in}
   Key Laboratory of Mathematics Mechanization, Institute of Systems Science, AMSS, Chinese Academy of Sciences, Beijing 100190, China\\
School of Mathematical Sciences, University of Chinese Academy of Sciences, Beijing 100049, China}

%\vspace{0.18in}

%\begin{tabular}{p{16cm}}
% \hline \\
%\end{tabular}

\vspace{0.18in}

\begin{abstract} \baselineskip=15pt \vspace{0.18in}
We investigate the initial-boundary value problem for the integrable spin-1 Gross-Pitaevskii (GP) equations with $4\times 4$ Lax pair on the half-line. The solution of this system can be obtained in terms of the solution of a $4\times 4$ matrix Riemann-Hilbert (RH) problem formulated in the complex $k$-plane. The relevant jump matrices of the RH problem can be explicitly found using the two spectral functions $s(k)$ and $S(k)$, which can be defined by the initial data, the Dirichlet-Neumann boundary data at $x=0$. The global relation is established between the two dependent spectral functions. The general mappings between Dirichlet and Neumann boundary values are analyzed in terms of the global relation.

%\vspace{0.1in} \noindent {\bf Keywords:}  Riemann-Hilbert problem; Initial-boundary value problem; half-line; Global relation; Spin-1 Gross-Pitaevskii equations

\end{abstract}

%\vspace{-0.05in}
%\begin{tabular}{p{16cm}}
 % \hline \\
%\end{tabular}

\vspace{-0.15in}

\maketitle

\baselineskip=15pt

\textbf{In 1967, Gardner, Greene, Kruskal, and Miura presented a powerful inverse scattering transformation (IST) to investigate solitons of the KdV equation with an initial value problem. After that this method was used to solve the initial value problems for many integrable nonlinear evolution partial differential equations (PDEs) with the Lax pairs. Moreover, the IST method was further extended such as the Fokas' unified transformation method. The Fokas unified method can be used to study the initial-boundary value problems for some integrable nonlinear integrable evolution PDEs with $2\times 2$ and $3\times 3$ Lax pairs on the half-line and the finite interval. To the best of our knowledge, so far there is no work on the IBV problems of integrable equations with $4\times 4$ Lax pairs on the half-line. In this paper, We investigate the initial-boundary value problem for the integrable spin-1 Gross-Pitaevskii (GP) equations with $4\times 4$ Lax pair on the half-line. The solution of this system can be obtained in terms of the solution of a $4\times 4$ matrix Riemann-Hilbert problem formulated in the complex $k$-plane. The relevant jump matrices of the RH problem can be explicitly found using the two spectral functions $s(k)$ and $S(k)$, which can be defined by the initial data, the Dirichlet-Neumann boundary data at $x=0$. The global relation is established between the two dependent spectral functions. The general mappings between Dirichlet and Neumann boundary values are analyzed in terms of the global relation. }

\section{Introduction}

\quad The initial value problems for many integrable nonlinear evolution partial differential equations (PDEs) with the Lax pairs
can be solved in terms of the inverse scattering transform (IST)~\cite{ist,soliton2, soliton}. After that, there exist some important extensions of the IST such as the Deift-Zhou nonlinear steepest descent method~\cite{rh} and the Fokas unified method~\cite{f1,f3,f4}. Particularly, the Fokas unified method can be used to
study the initial-boundary value problems for both linear and nonlinear integrable evolution PDEs with $2\times 2$ Lax pairs on the half-line and the finite interval, such as the nonlinear Schr\"odinger equation~\cite{f3,nls1,nls2,nls3,nls4}, the sine-Gordon equation~\cite{sg,sg2}, the KdV equation~\cite{kdv}, the mKdV equation~\cite{mkdv1,mkdv2}, the derivative nonlinear Schr\"odinger equation~\cite{dnls1}, Ernst equations~\cite{ernst}, and etc. (see Refs.~\cite{m1,m2,m3} and references therein).
Recently, Lenells extended the Fokas method to study the initial-boundary value (IBV) problems for integrable nonlinear evolution
equations with $3\times 3$ Lax pairs on the half-line~\cite{le12}. After that, the idea was extended to study IBV problems of
some integrable nonlinear evolution equations with $3\times 3$ Lax pairs on the half-line or the finite interval,
such as the Degasperis-Procesi equation~\cite{ds}, the Sasa-Satsuma equation~\cite{ss}, the coupled nonlinear Schr\"odinger equations~\cite{cnls1,cnls2,cnls3,cnls4}, and the Ostrovsky-Vakhnenko equation~\cite{os}. To the best of our knowledge, so far there is no work on the IBV problems of integrable equations with $4\times 4$ Lax pairs on the half-line.

The aim of this paper is to develop a methodology for analyzing the IBV problems for integrable nonlinear evolution equations with $4\times 4$
Lax pairs on the half-line by extending the method~\cite{f1,f3,f4,le12} for the integrable nonlinear PDEs with $2\times 2$ and $3\times 3$ Lax pairs. In this paper, we will study the IVB problem of the integrable spin-1 GP equations
\bee
\label{pnls}
\left\{\begin{array}{l}
\d i q_{1t}+ q_{1xx}-2\alpha\left(|q_1|^{2}+2|q_0|^2\right)q_1-2\alpha\beta q_0^2\bar{q}_{-1}=0, \vspace{0.1in}\\
\d i q_{0t}+ q_{0xx}-2\alpha\left(|q_1|^{2}+|q_0|^2+|q_{-1}|^2\right)q_0-2\alpha\beta q_1q_{-1}\bar{q}_0=0,  \vspace{0.1in}\\
\d i q_{-1t}+ q_{-1xx}-2\alpha\left(2|q_0|^{2}+|q_{-1}|^2\right)q_{-1}-2\alpha\beta q_0^2\bar{q}_1=0, \quad
\alpha^2=\beta^2=1,
 \end{array}\right.
 \ene
with the initial-boundary value conditions
 \bee\label{ibv}
 \begin{array}{lll}
 {\rm Initial\,\, conditions:} & q_j(x, t=0)=q_{0j}(x)\in \mathbb{S}(\mathbb{R}^+),&  j=1, 0, -1,\quad 0<x<\infty, \vspace{0.1in} \\
 {\rm Dirichlet \,\, boundary \,\, conditions:} & q_j(x=0, t)=u_{0j}(t), & j=1, 0, -1,\,\quad 0<t<T, \vspace{0.1in} \\
 {\rm Neumann \,\, boundary \,\, conditions:} &  q_{jx}(x=0, t)=u_{1j}(t),& j=1, 0, -1,\quad 0<t<T,
 \end{array} \ene
 where the complex-valued spinor condensate wave functions $q_j=q_j(x,t),\, j=1,0,-1$ are the sufficiently smooth functions defined in the finite region $\Omega=\left\{(x,t)\,|\, x\in [0, \infty),\, t\in [0, T]\right\}$ with $T>0$ being the fixed finite time, the overbar denotes the complex conjugate, $\mathbb{S}(\mathbb{R}^+)$ denotes the space of Schwartz functions,  the initial data $q_{0j}(x),\,j=1,0,-1$  and boundary data $u_{0j}(t), \, u_{1j}(t),\, j=1,0, -1$ are sufficiently smooth and compatible at points $(x,t)=(0, 0)$.

The spin-1 GP system (\ref{pnls}) can describe soliton dynamics of an $F=1$ spinor Bose-Einstein condensates~\cite{sgp}.
The four types of parameters: $(\alpha, \beta)=\{(1,1), (1, -1), (-1, 1), (-1, -1)\}$ in the spin-1 GP system (\ref{pnls})
 correspond to the four roles of the self-cross-phase modulation (nonlinearity) and spin-exchange modulation,  respectively, that is,
 (attractive, attractive), (attractive, repulsive), (repulsive, attractive), and ((repulsive, repulsive).
 In particular, Eq.~(\ref{pnls}) with the attractive mean¨Cfield nonlinearity and ferromagnetic spin-exchange modulation
 was shown to possess multi-bright soliton solutions~\cite{sgpb1}.  Eq.~(\ref{pnls}) with the repulsive mean-field nonlinearity and ferromagnetic spin-exchange modulation was shown to possess multi-dark soliton solutions~\cite{sgpd}. Moreover, double-periodic wave solutions of  Eq.~(\ref{pnls}) were also found~\cite{sgpp}.   System (\ref{pnls}) is associated with a variational principle
\bee
 iq_{jt}(x,t)=\frac{\delta \mathcal{E}_{GP}}{\delta \breve{q}_j(x,t)},\quad
  \breve{q}_1(x,t)=\bar{q}_1(x,t),\quad \breve{q}_0(x,t)=2\bar{q}_0(x,t),\quad \breve{q}_{-1}(x,t)=\bar{q}_{-1}(x,t),
 \ene
  with the energy functional being of the form
\bee\no
\mathcal{E}_{GP}=\d \int dx\Big\{\sum_{j=1,0,-1}|q_{jx}|^2+\alpha\left[|q_1|^4+|q_{-1}|^4+2|q_0|^4+4(|q_1|^2+|q_{-1}|^2)|q_0|^2\right]
  +2\alpha\beta\Re(q_0^2\bar{q}_1\bar{q}_{-1})\Big\}.
\ene

The rest of this paper is organized as follows. In Sec. 2, we study the  spectral analysis of the associated $4\times 4$ Lax pair of Eq.~(\ref{pnls}). Sec. 3 presents the corresponding $4\times 4$ matrix RH problem in terms of the jump matrices found in Sec. 2.
The global relation is used to establish the map between the Dirichlet and Neumann boundary values in Sec. 4.

\section{The spectral analysis of the Lax pair}

In this subsection, we will simultaneously consider the spectral analysis of the Lax pair (\ref{lax}) to
present sectionally its analytic eigenfunctions in order to formulate a $4\times 4$ matrix RH problem
defined in the complex $k$-plane.

\subsection*{(a) \, The closed one-form for the Lax pair}

\quad The spin-1 GP equations (\ref{pnls}) admits the $4 \times 4$ Lax pair~\cite{sgp}
\bee \label{lax}
\left\{\begin{array}{l}
                \psi_x+ik\sigma_4\psi=U(x,t)\psi,    \vspace{0.1in}  \\
                \psi_t+2ik^2\sigma_4\psi=V(x,t,k)\psi,
                 \end{array}\right.
    \ene
where $\psi=\psi(x,t,k)$ is a 4$\times$4 matrix-valued or $4\times 1$ column vector-valued spectral function, $k\in \mathbb{C}$ is an isospectral parameter, $\sigma_4={\rm diag}(1,1,-1,-1)$.  and the $4 \times 4$ matrix-valued functions $U(x,t)$ and $V(x,t,k)$ are defined by
\bee
U(x,t)=\left(\begin{array}{cccc}
            0 & 0 & q_1 & q_0 \vspace{0.05in}\\
            0&  0 & \beta q_0 & q_{-1} \vspace{0.05in}\\
            \alpha \bar{q}_1 & \alpha\beta \bar{q}_0 & 0 & 0 \vspace{0.05in}\\
            \alpha \bar{q}_0 & \alpha \bar{q}_{-1} & 0 & 0
            \end{array}\right), \quad V(x,t,k)=2kU+V_0,\quad V_0=i\sigma_4(U_{x}-U^2).
            \ene

A new eigenfunction $\mu=\mu(x,t,k)$ is defined by the transform
 \bee\label{mud}
 \mu(x,t,k)=\psi(x,t,k)e^{i(kx+2k^2t)\sigma_4},
 \ene
such that the Lax pair (\ref{lax}) is changed into an equivalent form
\bee\label{mulax}
    \left\{     \begin{array}{l}
                 \mu_x+ik\hat{\sigma}_4\mu= U(x,t)\mu,    \vspace{0.1in}            \\
                 \mu_t+2ik^2\hat{\sigma}_4\mu=V(x,t,k)\mu,
                 \end{array}\right.
    \ene
where $\hat{\sigma}_4\mu=[\sigma_4, \mu],$ $\hat{\sigma}_4$ denote the commutator with respect to $\sigma_4$ and the operator acting on a $4\times 4$ matrix $X$ by
$\hat{\sigma}_4X=[\sigma_4, X]$ such that $e^{x\hat{\sigma}_4}X=e^{x\sigma_4}Xe^{-x\sigma_4}$. The Lax pair (\ref{mulax}) leads to a full derivative form
\bee\label{dform}
d\left[e^{i(kx+2k^2t)\hat{\sigma}_4}\mu(x,t,k)\right]=W(x,t,k),
\ene
where the closed one-form $W(x,t,k)$ is
\bee \label{w}
 W(x,t,k)=e^{i(kx+2k^2t)\hat{\sigma}_4}[U(x,t)\mu(x,t,k)dx+V(x,t,k)\mu(x,t,k)dt].
 \ene

\subsection*{(b) \, The basic eigenfunctions $\mu_j's$ }

\begin{figure}[!t]
\begin{center}
{\scalebox{0.65}[0.65]{\includegraphics{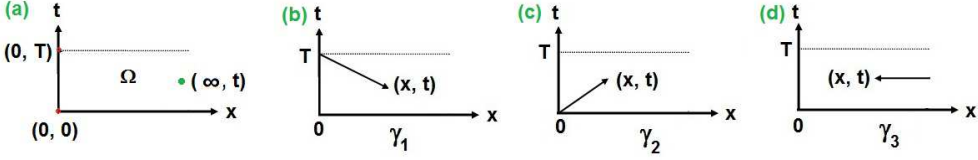}}}
\end{center}
\vspace{-0.3in}\caption{(a) the region $\Omega$; (b)-(d) three contours $\gamma_j \, (j=1,2,3)$ in the $(x,t)$-plane. }
\label{ga}
\end{figure}

\quad For any point $(x,t)$ in the considered region $\Omega=\{(x,t)| 0<x <\infty,\, 0< t< T\}$ (see Fig.~\ref{ga}(a)),  $\{\gamma_j\}_1^3$ denote
the three contours in the domain $\Omega$ connecting $(x_j, t_j)$ to $(x,t)$, respectively, where $(x_1, t_1)=(0, T),\quad  (x_2, t_2)=(0, 0),\quad  (x_3, t_3)=(\infty, t)$ (see Figs.~\ref{ga}(b)-(d)). Thus for the point $(\xi, \tau)$ on the each contour, we have
\bee\label{gammad}
 \begin{array}{rll}
 \gamma_1: & x-\xi \geq 0, & t-\tau \leq 0, \vspace{0.1in}\\
 \gamma_2: & x-\xi \geq 0, & t-\tau \geq 0, \vspace{0.1in}\\
 \gamma_3: & x-\xi \leq 0, & t-\tau = 0,
 \end{array}
\ene

It follows from the one-form (\ref{dform}) that we can use the Volterra integral equations to define its three eigenfunctions $\{\mu_j\}_1^3$ on the above-mentioned three contours $\{\gamma_j\}_1^3$
\bee \label{mus}
\begin{array}{rl}
\mu_j(x,t,k)=\d \mathbb{I}+\int_{(x_j, t_j)}^{(x,t)}e^{-i(kx+2k^2t)\hat{\sigma}_4}W_j(\xi,\tau,k),\quad j=1,2,3,\quad (x,t)\in \Omega,
\end{array}
\ene
where $\mathbb{I}={\rm diag}(1,1,1,1)$, the integral is over a piecewise smooth curve from $(x_j, t_j)$ to $(x,t)$,  $W_j(x,t,k)$ is given by Eq.~(\ref{w}) with $\mu(x,t,k)$ replaced by $\mu_j(x,t,k)$. Since the one-form $W_j$ is closed, thus $\mu_j$ is independent of the path of integration. If we choose the paths of integration to be parallel to the $x$ and $t$ axes, then the integral Eq.~(\ref{mus}) becomes ($j=12,3$)
\bee\label{musg}
\begin{array}{rl}
\mu_j=&\!\!\! \d \mathbb{I}+\int_{x_j}^x e^{-ik(x-\xi)\hat{\sigma}_4}(U\mu_j)(\xi,t,k)d\xi+e^{-ik(x-x_j)\hat{\sigma}_4}\int_{t_j}^te^{-2ik^2(t-\tau)\hat{\sigma}_4} (V\mu_j)(x_j,\tau,k)d\tau, \end{array}
\ene

 Eq.~(\ref{musg}) implies that the first, second, third, and fourth columns of the matrices $\mu_j(x,t,k)$'s contain these exponentials
\bes
\label{muc}
\bee
& [\mu_j]_1: \,\, e^{2ik(x-\xi)+4ik^2(t-\tau)},\quad e^{2ik(x-\xi)+4ik^2(t-\tau)},\quad \vspace{0.1in}\\
&[\mu_j]_2: \,\, e^{2ik(x-\xi)+4ik^2(t-\tau)},\quad e^{2ik(x-\xi)+4ik^2(t-\tau)},\quad \vspace{0.1in}\\
&[\mu_j]_3: \,\, e^{-2ik(x-\xi)-4ik^2(t-\tau)},\quad e^{-2ik(x-\xi)-4ik^2(t-\tau)}, \vspace{0.1in}\\
&[\mu_j]_4: \,\, e^{-2ik(x-\xi)-4ik^2(t-\tau)},\quad e^{-2ik(x-\xi)-4ik^2(t-\tau)},
\ene\ees

\begin{figure}[!t]
\begin{center}
{\scalebox{0.15}[0.15]{\includegraphics{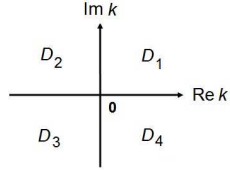}}}
\end{center}
\vspace{-0.25in}\caption{The domain $D_n\, (n=1,2,3,4)$ separating the complex $k$-plane. }
\label{kplane}
\end{figure}

To analyse the bounded domains of the eigenfunctions $\{\mu_j\}_1^3$ in the complex $k$-plane, we need to use the curve
$\mathbb{K}=\{k\in \mathbb{C} | \Re f(k) \cdot \Re g(k)=0,\, f(k)=ik,\, g(k)=ik^2\},$ to separate the complex $k$-plane into four regions (see Fig.~\ref{kplane}):
\bee
\begin{array}{l}
D_1=\{k\in\mathbb{C} \,|\, \Re f(k)=-\Im k<0 \,\, {\rm and} \,\,  \Re g(k)=-2\, \Re k\, \Im k<0\}, \vspace{0.1in} \\
D_2=\{k\in\mathbb{C} \,|\, \Re f(k)=-\Im k<0 \,\, {\rm and} \,\,  \Re g(k)=-2\, \Re k\, \Im k>0\}, \vspace{0.1in}\\
D_3=\{k\in\mathbb{C} \,|\,\Re f(k)=-\Im k>0 \,\, {\rm and} \,\, \Re g(k)=-2\, \Re k\, \Im k<0\}, \vspace{0.1in}\\
D_4=\{k\in\mathbb{C} \,|\, \Re f(k)=-\Im k>0\,\, {\rm and} \,\,  \Re g(k)=-2\, \Re k\, \Im k>0\},
\end{array}
\label{d}
\ene

Thus it follows from Eqs.~(\ref{gammad}), (\ref{muc}) and (\ref{d}) that the domains, where the different columns of eigenfunctions $\{\mu_j\}_1^3$ are bounded and analytic in the complex $k$-plane, are presented as follows:
\bee \label{muregion}
\left\{\begin{array}{l}
 \mu_1: (f_-(k) \cap g_+(k),\, f_-(k) \cap g_+(k),\, f_+(k) \cap g_-(k),\, f_+(k) \cap g_-(k))=: (D_2, D_2, D_3, D_3), \vspace{0.1in} \\
 \mu_2: (f_-(k) \cap g_-(k),\, f_-(k) \cap g_-(k),\, f_+(k) \cap g_+(k),\, f_+(k) \cap g_+(k))=: (D_1, D_1, D_4, D_4), \vspace{0.1in}\\
 \mu_3: (f_+(k) ,\, f_+(k) ,\, f_-(k) ,\, f_-(k))=: (C^-, C^-, C^+, C^+),
\end{array}\right.
\ene
where $C^-=D_3\cup D_4,\, C^+=D_1\cup D_2,\, f_+(k)=: \Re f(k)=-\Im k>0,\, f_-(k)=:\Re f(k)=-\Im k<0,\, g_+(k)=: \Re g(k)=-2\, \Re k\, \Im k>0,$  and $g_-(k)=:\Re g(k)=-2\, \Re k\, \Im k<0$.

\subsection*{(c) \, Symmetries of eigenfunctions}

For the convenience, we write a $4\times 4$ matrix $X=(X_{ij})_{4\times 4}$ as
\bee\label{mde}
\begin{array}{c}
X=\left(\begin{array}{cc} \tilde{X}_{11} & \tilde{X}_{12} \vspace{0.1in}\\ \tilde{X}_{21} & \tilde{X}_{22}\end{array} \right),\quad
\tilde{X}_{11}=\left(\begin{array}{cc} X_{11} & X_{12}  \vspace{0.1in}\\ X_{21} & X_{22} \end{array} \right),\quad
\tilde{X}_{12}=\left(\begin{array}{cc} X_{13} & X_{14}  \vspace{0.1in}\\ X_{23} & X_{24} \end{array} \right),\vspace{0.1in}\\
\tilde{X}_{21}=\left(\begin{array}{cc} X_{31} & X_{32}  \vspace{0.1in}\\ X_{41} & X_{42} \end{array} \right),\quad
\tilde{X}_{22}=\left(\begin{array}{cc} X_{33} & X_{34}  \vspace{0.1in}\\ X_{43} & X_{44} \end{array} \right),
\end{array}\ene

\quad
Let $\mathbb{U}(x,t, k)=-ik\sigma_4+U(x,t),\quad \mathbb{V}(x,t, k)=-2ik^2\sigma_4+V(x,t,k)$. Then the symmetry properties of $\mathbb{U}(x,t, k)$ and $\mathbb{V}(x,t, k)$ imply that the eigenfunction $\mu(x,t,k)$ have the symmetries
\bee
(\tilde{\mu}(x,t,k))_{11}=P^{\beta}\,\overline{(\tilde{\mu}(x,t,\bar{k}))}_{22}P^{\beta},\quad
(\tilde{\mu}(x,t,k))_{12}=\alpha\,\overline{(\tilde{\mu}(x,t,\bar{k}))}_{21}^T,
\ene
where $P^{\beta}={\rm diag}(1, \beta),\, \beta^2=1$.

Since
\bee\no
P_{\pm}^{\alpha}\,\overline{\mathbb{U}(x,t, \bar{k})}P_{\pm}^{\alpha}=-\mathbb{U}(x,t,k)^T, \quad
P_{\pm}^{\alpha}\,\overline{\mathbb{V}(x,t, \bar{k})}P_{\pm}^{\alpha}=-\mathbb{V}(x,t,k)^T,
\ene
where $P_{\pm}^{\alpha}={\rm diag}(\pm\alpha, \pm\alpha,  \mp 1, \mp 1),\, \alpha^2=1$.

According to Eq.~(\ref{mualax}) (see the similar proof in Ref.~\cite{nls3}), we know that
the eigenfunction $\psi(x,t,k)$ of the Lax pair (\ref{lax}) and $\mu(x,t,k)$ of the Lax pair (\ref{mulax}) are  of
the same symmetric relation
\bee\label{symmetry}
\begin{array}{l}
\psi^{-1}(x,t,k)=P_{\pm}^{\alpha}\,\overline{\psi(x,t,\bar{k})}^TP_{\pm}^{\alpha},\quad \mu^{-1}(x,t,k)=P_{\pm}^{\alpha}\,\overline{\mu(x,t,\bar{k})}^TP_{\pm}^{\alpha},
\end{array}
\ene

Moreover, In the domains where $\mu$ is bounded, we have
\bee
 \mu(x,t,k)=\mathbb{I}+O\left(\frac{1}{k}\right),\quad k\to \infty,
\ene
and ${\rm det} [\mu(x,t,k)]=1$ since ${\rm tr} (\mathbb{U}(x,t, k))={\rm tr} (\mathbb{V}(x,t,k))=0$.

\subsection*{(d) \,  The minors of eigenfunctions}

\quad The cofactor matrix $X^A$ (or the transpose of the adjugate) of a $4\times 4$ matrix $X$ is given by
\bee
{\rm adj}(X)^T=X^A=\left(\begin{array}{rrrr}
 m_{11}(X) & -m_{12}(X) &  m_{13}(X) &  -m_{14}(X) \vspace{0.05in}\\
 -m_{21}(X) & m_{22}(X) &  -m_{23}(X) &  m_{24}(X) \vspace{0.05in}\\
 m_{31}(X) & -m_{32}(X) &  m_{33}(X) &  -m_{34}(X) \vspace{0.05in}\\
 -m_{41}(X) & m_{42}(X) &  -m_{43}(X) &  m_{44}(X)
\end{array}\right),
\ene
where $m_{ij}(X)$ denote the $(ij)$th minor of $X$ and $(X^A)^TX ={\rm adj}(X) X=\det X$.

It follows from Eq.~(\ref{mulax}) that be shown that the matrix-valued functions $\mu_j^A$'s satisfy the Lax pair
\bee\label{mualax}
    \left\{   \begin{array}{l}
                 \mu_{j,x}^A-ik\hat{\sigma}_4\mu_j^A= -U^T\mu_j^A,    \vspace{0.1in} \\
                 \mu_{j,t}^A-2ik^2\hat{\sigma}_4\mu_j^A=-V^T\mu_j^A,
                 \end{array} \right.
                 \ene
whose solutions can be expressed as
\bee\begin{array}{rl}
\mu_j^A(x,t,k)=&\!\!\! \d \mathbb{I}-\int_{x_j}^x e^{ik(x-\xi)\hat{\sigma}_4}(U\mu_j^A)(\xi,t,k)d\xi  -e^{ik(x-x_j)\hat{\sigma}_4}\int_{t_j}^te^{2ik^2(t-\tau)\hat{\sigma}_4} (V\mu_j^A)(x_j,\tau,k)d\tau,
\end{array}
\ene
by using the Volterra integral equations, where $U^T$ and $V^T$ denote the  transposes of $U$ and $V$, respectively.

It is easy to check that the regions of boundedness of $\mu_j^A$:
\bee \no \left\{\begin{array}{l}
 \mu_1^A(x,t,k) {\rm \,\, is \,\, bounded\,\, for\,\,} k\in (D_3, D_3, D_2, D_2), \vspace{0.1in}\\
 \mu_2^A(x,t,k) {\rm \,\, is \,\, bounded\,\, for\,\,} k\in (D_4, D_4, D_1, D_1), \vspace{0.1in}\\
 \mu_3^A(x,t,k) {\rm \,\, is \,\, bounded\,\, for\,\,} k\in (C^+, C^+, C^-, C^-),
\end{array}\right.
\ene
which are symmetric ones of $\mu_j$ about the $\Re k$-axis (cf. Eq.~(\ref{muregion})).

\subsection*{(e) \, The spectral functions and the global relation}

\begin{figure}[!t]
\begin{center}
{\scalebox{0.2}[0.2]{\includegraphics{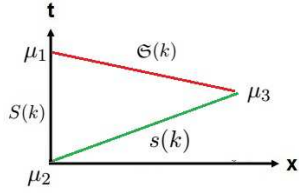}}}
\end{center}
\vspace{-0.3in}\caption{The relations among the dependent eigenfunctions $\mu_j(x,t,k),\, j=1,2,3$. }
\label{mu}
\end{figure}

\quad Let us introduce the $4\times 4$ matrix-valued functions $S(k),\, s(k)$, and $\mathfrak{S}(k)$ by $\mu_j,\, j=1,2,3$
\bee\label{mur}\left\{\begin{array}{l}
\mu_1(x,t,k)=\mu_2(x,t,k)e^{-i(kx+2k^2t)\hat{\sigma}_4}S(k), \vspace{0.1in}\\
\mu_3(x,t,k)=\mu_2(x,t,k)e^{-i(kx+2k^2t)\hat{\sigma}_4}s(k), \vspace{0.1in}\\
\mu_3(x,t,k)=\mu_1(x,t,k)e^{-i(kx+2k^2t)\hat{\sigma}_4}\mathfrak{S}(k),
\end{array} \right.
\ene

Evaluating  system (\ref{mur}) at $(x,t)=(0,0)$ and $(x, t)=(0, T)$, respectively, we have
\bee\label{sss}
\left\{\begin{array}{l}
 S(k)=\mu_1(0,0,k)=e^{2ik^2T\hat{\sigma}_4}\mu_2^{-1}(0,T,k),\vspace{0.1in}\\
 s(k)=\mu_3(0,0,k),\vspace{0.1in}\\
 \mathfrak{S}(k)=\mu_1^{-1}(0,0,k)\mu_3(0,0,k)=S^{-1}(k)s(k)=e^{2ik^2T\hat{\sigma}_4}\mu_3(0,T,k),
\end{array} \right.
\ene
These relations among $\mu_j$ are displayed in Fig.~\ref{mu}. Thus these three functions $S(k),\, s(k)$, and $\mathfrak{S}(k)$ are dependent such that we only consider two of them, e.g., $S(k)$ and $s(k)$.

According to the definition (\ref{musg}) of $\mu_j$, Eq.~(\ref{sss})  implies that
\bee\label{muss}
\begin{array}{rl}
s(k)=&\!\!\!\d \mathbb{I}-\int_{0}^{\infty} e^{ik\xi\hat{\sigma}_4}(U\mu_3)(\xi,0,k)d\xi \vspace{0.1in}\\
S(k)=&\!\!\! \d \mathbb{I}-\int_0^T e^{2ik^2\tau\hat{\sigma}_4}(V\mu_1)(0, \tau,k)d\xi =\left[\mathbb{I}+\int_0^T e^{2ik^2\tau\hat{\sigma}_4}(V\mu_2)(0,\tau,k)d\tau\right]^{-1},
  \end{array}
\ene
where $\mu_{j}(0,t,k),\ j=1,2 $ and $\mu_3(x,0,k),\, 0<x<\infty,\, 0<t<T$ satisfy the Volterra integral equations
\bee \label{mu123}
\begin{array}{l}
\mu_3(x,0,k)=\d \mathbb{I}-\int_x^{\infty}e^{-ik(x-\xi)\hat{\sigma}_4} (U\mu_3)(\xi,0,k)d\xi,\quad 0<x<\infty,\,\,
k\in (C^-, C^-, C^+, C^+), \vspace{0.1in}\\
\mu_1(0,t,k)=\d \mathbb{I}-\int_t^Te^{-2ik^2(t-\tau)\hat{\sigma}_4} (V\mu_1)(0,\tau,k)d\tau,\quad 0<t<T, \quad  k\in (D_2\cup U_4, D_2\cup U_4, D_1\cup U_3, D_1\cup U_3),
\vspace{0.1in}\\
\mu_2(0,t,k)=\d \mathbb{I}+\int_0^te^{-2ik^2(t-\tau)\hat{\sigma}_4} (V\mu_2)(0,\tau,k)d\tau, \quad 0<t<T,\quad  k\in (D_1\cup U_3, D_1\cup U_3, D_2\cup U_4, D_2\cup U_4),
\end{array}
\ene
Thus, it follows from Eqs.~(\ref{muss}) and (\ref{mu123}) that $s(k)$ and $S(k)$ are determined by $U(x,0,k)$ and $V(0,t,k)$, i.e., by the initial data $q_{j}(x, t=0)$ and the Dirichlet-Neumann boundary data $q_{j}(x=0, t)$ and $q_{jx}(x=0, t),\, j=1,0,-1$, respectively. In fact, $\mu_3(x,0,k)$
and $\mu_{1,2}(0,t,k)$ satisfy the $x$-part and $t$-part of the Lax pair (\ref{mulax}) at $t=0$ and $x=0$, respectively, that is,
\bee
x-{\rm part}:\, \left\{\begin{array}{l}
\mu_{x}(x,0,k)+ik[\sigma_4, \mu(x,0,k)]=U(x, t=0)\mu(x,0,k), \vspace{0.1in}\\
\d\lim_{x\to\infty}\mu(x,0,k)=\mathbb{I},\quad 0<x<\infty,
\end{array}\right.
\ene
\bee
t-{\rm part}:\, \left\{\begin{array}{l}
\mu_{t}(0,t,k)+2ik^2[\sigma_4, \mu(0,t,k)]=V(x=0,t,k)\mu(0,t,k), \,\, 0<t<T,\vspace{0.1in}\\
\mu(0,0,k)=\mathbb{I}, \quad \mu(0,T,k)=\mathbb{I},
\end{array}\right.
\ene

Moreover, the functions  $\{S(k),\, s(k)\}$ and  $\{S^A(k),\, s^A(k)\}$ have the following boundedness:
 \bee\no
\left\{\begin{array}{l}
 S(k){\rm \,\, is \,\, bounded\,\, for\,\,} k\in  (D_2\cup D_4, D_2\cup D_4, D_1\cup D_3, D_1\cup D_3), \vspace{0.1in} \\
 s(k){\rm \,\, is \,\, bounded\,\, for\,\,} k\in  (C^-, C^-, C^+, C^+), \vspace{0.1in}\\
 S^A(k){\rm \,\, is \,\, bounded\,\, for\,\,} k\in  (D_1\cup D_3, D_1\cup D_3, D_2\cup D_4, D_21\cup D_4), \vspace{0.1in} \\
 s^A(k){\rm \,\, is \,\, bounded\,\, for\,\,} k\in  (C^+, C^+, C^-, C^-),
\end{array} \right.
\ene

It follows from the third one in  Eq.~(\ref{sss}) that we have the so-called global relation
\bee\label{srr2}
c(T,k)=\mu_3(0, T, k)=e^{-2ik^2T\hat{\sigma}_4}[S^{-1}(k)s(k)],
\ene
where $\mu_3(0,t,k),\, 0<t<T$ satisfies the Volterra integral equation
\bee
\mu_3(0,t,k)=\d \mathbb{I}-\int_{0}^{\infty} e^{ik\xi\hat{\sigma}_4}(U\mu_3)(\xi,t,k)d\xi,\quad 0<t<T, \quad
 k\in (C^-, C^-, C^+, C^+),
\ene

\subsection*{(f)  \, The definition of matrix-valued functions $M_n$'s}

\quad In each domain $D_n, \, n=1,2,3,4$ of the complex $k$-plane,  the solution $M_n(x,t,k)$ of Eq.~(\ref{mulax}) is
\bee \label{mn}
(M_n(x,t,k))_{lj}=\delta_{lj}+\int_{(\gamma^n)_{sj}}\left(e^{-i(kx+2k^2t)\hat{\sigma}_4}W_n(\xi,\tau,k)\right)_{lj}, \quad k\in D_n,\quad  l,j=1,2,3,4.
\ene
via the Volterra integral equations, where $W_n(x,t,k)$ is given by Eq.~(\ref{w}) with $\mu(x,t,k)$ replaced with $M_n(x,t,k)$, and the definition of the contours $(\gamma^n)_{lj}$'s is given by
\bee\label{gamma}
(\gamma^n)_{lj}=\left\{ \begin{array}{l}
\gamma_1,\,\,\, {\rm if} \,\,  \Re f_l(k)< \Re f_j(k) \,\,  {\rm and} \,\, \Re g_l(k)\geq \Re g_j(k), \vspace{0.1in} \\
\gamma_2,\,\,\, {\rm if}  \,\, \Re f_l(k)< \Re f_j(k) \,\, {\rm and} \,\, \Re g_l(k) < \Re g_j(k), \vspace{0.1in}\\
\gamma_ 3,\,\,\, {\rm if} \,\,  \Re f_l(k) \geq \Re f_j(k),
\end{array}\right.
\ene
for $k\in D_n$, where $f_{1,2}(k)=-f_{3,4}(k)=-ik,\, g_{1,2}(k)=-g_{3,4}(k)=-ik^2$.

The definition (\ref{gamma}) of $(\gamma^n)_{lj}$  implies that the matrices $\gamma^n\, (n=1,2,3,4)$ are of the forms
\bee\begin{array}{r}
\gamma^1=\left(
\begin{array}{cccc}
 \gamma_3 & \gamma_3 & \gamma_3 & \gamma_3 \\
 \gamma_3 & \gamma_3 & \gamma_3 & \gamma_3 \\
 \gamma_2 & \gamma_2 & \gamma_3 & \gamma_3 \\
 \gamma_2 & \gamma_2 & \gamma_3 & \gamma_3
 \end{array}
\right),
\gamma^2=\left(
\begin{array}{cccc}
 \gamma_3 & \gamma_3 & \gamma_3 & \gamma_3 \\
 \gamma_3 & \gamma_3 & \gamma_3 & \gamma_3 \\
 \gamma_1 & \gamma_1 & \gamma_3 & \gamma_3 \\
 \gamma_1 & \gamma_1 & \gamma_3 & \gamma_3
 \end{array}
\right),
\gamma^3=\left(
\begin{array}{cccc}
 \gamma_3 & \gamma_3 & \gamma_1 & \gamma_1 \\
 \gamma_3 & \gamma_3 & \gamma_1 & \gamma_1 \\
 \gamma_3 & \gamma_3 & \gamma_3 & \gamma_3 \\
 \gamma_3 & \gamma_3 & \gamma_3 & \gamma_3
 \end{array}
\right),
\gamma^4=\left(
\begin{array}{cccc}
 \gamma_3 & \gamma_3 & \gamma_2 & \gamma_2 \\
 \gamma_3 & \gamma_3 & \gamma_2 & \gamma_2 \\
 \gamma_3 & \gamma_3 & \gamma_3 & \gamma_3 \\
 \gamma_3 & \gamma_3 & \gamma_3 & \gamma_3
 \end{array}
\right),
\end{array}\ene

According to the similar proof for the $3\times 3$ Lax pair in ~\cite{le12} and the above-mentioned properties of $\mu(x,t,k)$,
we have the bounedness and analyticity of $M_n$:

\vspace{0.1in}
\noindent {\bf Proposition 2.1.} {\it The matrix-valued functions $M_n(x,t,k),\, n=1,2,3,4$ are weill defined by Eq.~(\ref{mn}) for $k\in \bar{D}_n$ and $(x,t)\in \bar{\Omega}$. For any fixed point $(x,t)$, $M_n$'s are the bounded and analytic function of $k\in D_n$ away from a possible discrete set of singularity $\{k_j\}$ at which the Fredholm determinants vanish. $M_n(x,t,k)$ also admits the bounded and continuous extensions to
$\bar{D}_n$ and
\bee\label{ml}
 M_n(x,t,k)=\mathbb{I}+O\left(\frac{1}{k}\right),\,\,k\in D_n,\,\ k\to\infty,\,\,  n=1,2,3,4.
\ene}

\subsection*{(g) \, The jump matrices}

\quad The new spectral functions $S_n(k)\, (n=1,2,3, 4)$ are introduced by
\bee
 S_n(k)=M_n(0,0,k), \quad k\in D_n,\quad n=1,2,3,4.
\ene
Let $M(x,t,k)$ stand for the sectionally analytic function on the Riemann $k$-spere which is equivalent to $M_n(x,t,k)$ for $k\in D_n$. Then $M(x,t,k)$ solves the jump equations
\bee\label{jumpc}
 M_n(x,t,k)=M_m(x,t,k)J_{mn}(x,t,k),\quad k\in \bar{D}_n\cap \bar{D}_m, \quad n,m=1,2,3,4,\quad n\not= m,
\ene
with the jump matrices $J_{mn}(x,t,k)$ defined by
\bee\label{jump}
J_{mn}(x,t,k)=e^{-i(kx+2k^2t)\hat{\sigma}_4}[S_m^{-1}(k)S_n(k)].
\ene

\vspace{0.1in}
\noindent {\bf Proposition 2.2.} {\it The matrix-valued functions $S_n(x,t,k)\, (n=1,2,3,4)$ defined by
\bee\label{mns}
 M_n(x,t,k)=\mu_2(x,t,k)e^{-i(kx+2k^2t)\hat{\sigma}_4}S_n(k), \quad k\in D_n,
 \ene
can be determined by the entries of $s(k)=(s_{ij})_{4\times 4},\, S(k)=(S_{ij})_{4\times 4}$ (cf. Eq.~(\ref{sss})) as follows:
\bee\label{sn}
\begin{array}{ll}
S_1(k)=\left(\begin{array}{cccc}
 \frac{m_{22}(s)}{n_{33,44}(s)} &  \frac{m_{21}(s)}{n_{33,44}(s)} &  s_{13} & s_{14} \vspace{0.1in}\\
 \frac{m_{12}(s)}{n_{33,44}(s)} &  \frac{m_{11}(s)}{n_{33,44}(s)} &  s_{23} & s_{24} \vspace{0.1in}\\
 0 &  0 &  s_{33} & s_{34} \vspace{0.1in}\\
 0 &  0 &  s_{43} & s_{44}
 \end{array}
\right), \quad
S_2(k)=\left(\begin{array}{cccc}
 S_2^{(11)} &  S_2^{(12)} & s_{13} & s_{14} \vspace{0.1in}\\
 S_2^{(21)} &  S_2^{(22)} &  s_{23} & s_{24} \vspace{0.1in}\\
  S_2^{(31)} &  S_2^{(32)}  &  s_{33} & s_{34} \vspace{0.1in}\\
  S_2^{(41)} &  S_2^{(42)} &  s_{43} & s_{44}
 \end{array}
\right), \vspace{0.15in}\\
S_3(k)=\left(\begin{array}{cccc}
 s_{11} &  s_{12} & S_3^{(13)} & S_3^{(14)} \vspace{0.1in}\\
 s_{21} &  s_{22} &  S_3^{(23)} & S_3^{(24)} \vspace{0.1in}\\
 s_{31} &  s_{32}  & S_3^{(33)} & S_3^{(34)} \vspace{0.1in}\\
 s_{41} &  s_{42} & S_3^{(43)} & S_3^{(44)}
 \end{array}
\right),\quad
S_4(k)=\left(\begin{array}{cccc}
 s_{11} & s_{12} & 0  & 0 \\
 s_{21} & s_{22} & 0 &  0 \\
 s_{31} & s_{32} & \frac{m_{44}(s)}{n_{11,22}(s)} &  \frac{m_{43}(s)}{n_{11,22}(s)} \vspace{0.1in}\\
s_{41} & s_{42} & \frac{m_{34}(s)}{n_{11,22}(s)} &  \frac{m_{33}(s)}{n_{11,22}(s)}    \end{array}
\right),
\end{array}
\ene
where $n_{i_1j_1,i_2j_2}(X)$ denotes the determinant of the sub-matrix generated by taking the cross elements of $i_{1,2}$th rows and $j_{1,2}$th columns of the $4\times 4$ matrix $X$
and
\bee\no
\left\{\begin{array}{l}
S_2^{(1j)}=\dfrac{n_{1j,2(3-j)}(S)m_{2(3-j)}(s)+n_{1j,3(3-j)}(S)m_{3(3-j)}(s)+n_{1j,4(3-j)}(S)m_{4(3-j)}(s)}{\mathcal{N}([S]_1[S]_2[s]_3[s]_4)}, \vspace{0.1in}\\
S_2^{(2j)}=\dfrac{n_{2j,1(3-j)}(S)m_{1(3-j)}(s)+n_{2j,3(3-j)}(S)m_{3(3-j)}(s)+n_{2j,4(3-j)}(S)m_{4(3-j)}(s)}{\mathcal{N}([S]_1[S]_2[s]_3[s]_4)}, \vspace{0.1in}\\
S_2^{(3j)}=\dfrac{n_{3j,1(3-j)}(S)m_{1(3-j)}(s)+n_{3j,2(3-j)}(S)m_{2(3-j)}(s)+n_{3j,4(3-j)}(S)m_{4(3-j)}(s)}{\mathcal{N}([S]_1[S]_2[s]_3[s]_4)}, \vspace{0.1in}\\
S_2^{(4j)}=\dfrac{n_{4j,1(3-j)}(S)m_{1(3-j)}(s)+n_{4j,2(3-j)}(S)m_{2(3-j)}(s)+n_{4j,3(3-j)}(S)m_{3(3-j)}(s)}{\mathcal{N}([S]_1[S]_2[s]_3[s]_4)},
\end{array}\right.\quad j=1,2,
\ene
\bee\no\left\{\begin{array}{l}
S_3^{(1j)}=\dfrac{n_{1j,2(7-j)}(S)m_{2(7-j)}(s)+n_{1j,3(7-j)}(S)m_{3(7-j)}(s)+n_{1j,4(7-j)}(S)m_{4(7-j)}(s)}{\mathcal{N}([s]_1[s]_2[S]_3[S]_4)},\vspace{0.1in}\\
S_3^{(2j)}=\dfrac{n_{2j,1(7-j)}(S)m_{1(7-j)}(s)+n_{2j,3(7-j)}(S)m_{3(7-j)}(s)+n_{2j,4(7-j)}(S)m_{4(7-j)}(s)}{\mathcal{N}([s]_1[s]_2[S]_3[S]_4)}, \vspace{0.1in}\\
S_3^{(3j)}=\dfrac{n_{3j,1(7-j)}(S)m_{1(7-j)}(s)+n_{3j,2(7-j)}(S)m_{2(7-j)}(s)+n_{3j,4(7-j)}(S)m_{4(7-j)}(s)}{\mathcal{N}([s]_1[s]_2[S]_3[S]_4)}, \vspace{0.1in}\\
S_3^{(4j)}=\dfrac{n_{4j,1(7-j)}(S)m_{1(7-j)}(s)+n_{4j,2(7-j)}(S)m_{2(7-j)}(s)+n_{4j,3(7-j)}(S)m_{3(7-j)}(s)}{\mathcal{N}([s]_1[s]_2[S]_3[S]_4)},
\end{array}\right. \quad j=3,4,
\ene
where $\mathcal{N}([S]_1[S]_2[s]_3[s]_4)={\rm det}(n([S]_1, [S]_2, [s]_3, [s]_4))$ denotes the determinant of the matrix generated by choosing the first and second columns of $S(k)$ and the third and fourth columns of $s(k)$, and $\mathcal{N}([s]_1[s]_2[S]_3[S]_4)={\rm det}(n([s]_1, [s]_2, [S]_3, [S]_4))$. }

\vspace{0.1in}
\noindent {\bf Proof.}\, Let $\gamma_3^{x_0}$ with $x_0>0$ denote the contour $(x_0, 0)\to (x,t)$ in the $(x,t)$-plane
and $\mu_3(x,t,k; x_0)$ be determined by Eq.~(\ref{mus}) with $j=3$ and the contour $\gamma_3$ replaced by $\gamma_3^{x_0}$.
$M_n(x,t,k; x_0)$ is defined by Eq.~(\ref{mn}) with  the contour $\gamma_3$ replaced by $\gamma_3^{x_0}$.

 We introduce the functions $R_n(k; x_0), S_n(k; x_0)$, and $T_n(k; x_0)$  in the form
\bee\label{rstp}
\left\{\begin{array}{l}
M_n(x,t,k; x_0)=\mu_1(x,t,k)e^{-i(kx+2k^2t)\hat{\sigma}_4}R_n(k; x_0), \vspace{0.1in}\\
M_n(x,t,k; x_0)=\mu_2(x,t,k)e^{-i(kx+2k^2t)\hat{\sigma}_4}S_n(k; x_0), \vspace{0.1in}\\
M_n(x,t,k; x_0)=\mu_3(x,t,k; x_0)e^{-i(kx+2k^2t)\hat{\sigma}_4}T_n(k; x_0),
\end{array}\right.
\ene

It follows from Eq.~(\ref{rstp}) that we have the relations
\bee
\left\{\begin{array}{l}
R_n(k; x_0)=e^{2ik^2T\hat{\sigma}_4}M_n(0,T,k; x_0),\vspace{0.1in}\\
S_n(k; x_0)=M_n(0,0,k; x_0),\vspace{0.1in}\\
T_n(k; x_0)=e^{ikx_0\hat{\sigma}_4}[\mu_3^{-1}(x_0,0,k;x_0)M_n(x_0,0,k; x_0)],
\end{array}\right.
\ene
and
\bee \label{sneq}
\left\{\begin{array}{l}
S(k)=\mu_1(0,0,k)=S_n(k; x_0)R_n^{-1}(k; x_0), \vspace{0.1in}\\
s(k; x_0)=\mu_3(0,0,k; x_0)=S_n(k; x_0)T_n^{-1}(k; x_0),
\end{array}\right.
\ene
which can in general deduce the functions $\{R_n(k; x_0), S_n(k; x_0), T_n(k; x_0)\}$ for the given spectral functions $\{s(k), S(k)\}$.

Moreover, we can also determine some entries of $\{R_n(k; x_0), S_n(k; x_0), T_n(k; x_0)\}$ as
\bee
\left\{
\begin{array}{l}
(R_n(k; x_0))_{ij}=0, \,\,\,\,\, {\rm if} \,\,\, (\gamma^n)_{ij}=\gamma_1, \vspace{0.1in} \\
(S_n(k; x_0))_{ij}=0,  \,\,\,\,\, {\rm if} \,\,\,  (\gamma^n)_{ij}=\gamma_2, \vspace{0.1in}\\
(T_n(k; x_0))_{ij}=\delta_{ij}, \,\,\, {\rm if} \,\,\,  (\gamma^n)_{ij}=\gamma_3,
\end{array}
\right.
 \ene
by using Eqs.~(\ref{mn}) and (\ref{rstp}). System (\ref{sneq}) contains $32$ scalar equations for $32$ unknowns.
Thus it follows from system (\ref{sneq}) that we have $S_n(k; x_0)$. Then taking the limit $x_0\to\infty$ of $S_n(k; x_0)$ yields Eq.~(\ref{sn}). $\square$

\subsection*{(h) \, The residue conditions}

\quad Since $\mu_2(x,t,k)$ is an entire function, it follows from Eq.~(\ref{mns}) that $M(x,t,k)$ only has the singularities at the points where
the $S_n(k)$'s have the singularities. The $S_n(k)$'s given by  Eq.~(\ref{sn}) imply that the possible singularities of $M(x,t,k)$ are as follows:

\begin{itemize}

\item {} $[M]_j,\ j=1,2$ could have poles in $D_1$ at the zeros of $n_{33,44}(s)(k)$;
\item {} $[M]_j,\, j=1,2$ could have poles in $D_2$ at the zeros of $\mathcal{N}([S]_1[S]_2[s]_3[s]_4)(k)$;
\item {} $[M]_j,\, j=3,4$ could have poles in $D_3$ at the zeros of $\mathcal{N}([s]_1[s]_2[S]_3[S]_4)(k)$;
\item {} $[M]_j,\, j=3,4$ could have poles in $D_4$ at the zeros of $n_{11,22}(s)(k)$.
\end{itemize}

We use $\{k_j\}_1^N$ to denote the above-mentioned possible zeros and suppose that they satisfy the following assumption.

\vspace{0.1in}
\noindent {\bf Assumption 2.3.} {\it We suppose  that
\begin{itemize}

\item {} $n_{33,44}(s)(k)$ admits $n_1$ possible simple zeros in $D_1$ denoted by $\{k_j\}_1^{n_1}$;

\item {} $\mathcal{N}([S]_1[S]_2[s]_3[s]_4)(k)$ admits $n_2-n_1$ possible simple zeros in $D_2$ denoted by $\{k_j\}_{n_1+1}^{n_2}$;

\item {} $\mathcal{N}([s]_1[s]_2[S]_3[S]_4)(k)$ admits $n_3-n_2$ possible simple zeros in $D_3$ denoted by $\{k_j\}_{n_2+1}^{n_3}$;

\item {} $n_{11,22}(s)(k)$ admits $N-n_3$ possible simple zeros in $D_4$ denoted by $\{k_j\}_{n_3+1}^N$;
\end{itemize}
and that none of these simple zeros coincide. Moreover, none of these functions are assumed to have zeros on the boundaries of the $D_n$'s ($n=1,2,3,4)$.}

\vspace{0.1in}
\noindent {\bf Proposition 2.4. } {\it Let $\{M_n(x,t,k)\}_1^4$ be the eigenfunctions given by Eq.~(\ref{mn}) and suppose that the set $\{k_j\}_1^N$ of singularities are as the above-mentioned Assumption 2.3. Then we have the following residue conditions:
\bee
 \label{rm1a} \begin{array}{rl}
\d{\rm Res}_{k=k_j}[M_1(x,t,k)]_l=&\!\! \dfrac{m_{2(3-l)}(s)(k_j)s_{24}(k_j)-m_{1(3-l)}(s)(k_j)s_{14}(k_j)}
            {\dot{n}_{33,44}(s)(k_j)n_{13,24}(s)(k_j)}[M_1(x,t,k_j)]_3e^{-2\theta(k_j)} \vspace{0.1in}\\
        &    +\dfrac{m_{1(3-l)}(s)(k_j)s_{13}(k_j)-m_{2(3-l)}(s)(k_j)s_{23}(k_j)}
          {\dot{n}_{33,44}(s)(k_j)n_{13,24}(s)(k_j)}[M_1(x,t,k_j)]_4e^{-2\theta(k_j)},  \vspace{0.1in}\\
        & \quad {\rm for}\quad 1\leq j\leq n_1,\quad k\in D_1,\quad l=1,2,
        \end{array}
\ene
\bee
 \label{rm2a} \begin{array}{rl}
 {\rm Res}_{k=k_j}[M_2(x,t,k)]_l=&\!\! \dfrac{S_{2}^{(1l)}(k_j)s_{24}(k_j)-S_2^{(2l)}(k_j)s_{14}(k_j)}
          {\dot{\mathcal{N}}([S]_1[S]_2[s]_3[s]_4)(k_j)n_{13,24}(s)(k_j)}[M_2(x,t,k_j)]_3e^{-2\theta(k_j)} \vspace{0.1in}\\
       &   +\dfrac{S_2^{(2l)}(k_j)s_{13}(k_j)-S_2^{(1l)}(k_j)s_{23}(k_j)}
          {\dot{\mathcal{N}}([S]_1[S]_2[s]_3[s]_4)(k_j)n_{13,24}(s)(k_j)}[M_2(x,t,k_j)]_4e^{-2\theta(k_j)}, \vspace{0.1in}\\
        & \quad {\rm for}\quad n_1+1\leq j\leq n_2,\quad k\in D_2, \quad l=1,2,
   \end{array}
   \ene
\bee
\label{rm3a} \begin{array}{rl}
 {\rm Res}_{k=k_j}[M_3(x,t,k)]_l=&\!\! \dfrac{S_{3}^{(1l)}(k_j)s_{22}(k_j)-S_3^{(2l)}(k_j)s_{12}(k_j)}
          {\dot{\mathcal{N}}([s]_1[s]_2[S]_3[S]_4)(k_j)n_{11,22}(s)(k_j)}M_3(x,t,k_j)]_1e^{2\theta(k_j)} \vspace{0.1in}\\
      &    +\dfrac{S_3^{(2l)}(k_j)s_{11}(k_j)-S_3^{(1l)}(k_j)s_{21}(k_j)}
          {\dot{\mathcal{N}}([s]_1[s]_2[S]_3[S]_4)(k_j)n_{11,22}(s)(k_j)}[M_3(x,t,k_j)]_2e^{2\theta(k_j)}, \vspace{0.1in}\\
      & \quad {\rm for}\quad n_2+1\leq j\leq n_3,\quad k\in D_3, \quad l=3,4,
   \end{array}
   \ene
\bee
 \label{rm4a} \begin{array}{rl}
 {\rm Res}_{k=k_j}[M_4(x,t,k)]_l=&\!\! \dfrac{m_{4(7-l)}(s)(k_j)s_{42}(k_j)-m_{3(7-l)}(s)(k_j)s_{32}(k_j)}
          {\dot{n}_{11,22}(s)(k_j)n_{31,42}(s)(k_j)}[M_4(x,t,k_j)]_1e^{2\theta(k_j)} \vspace{0.1in}\\
        &  +\dfrac{m_{3(7-l)}(s)(k_j)s_{31}(k_j)-m_{4(7-l)}(s)(k_j)s_{41}(k_j)}
          {\dot{n}_{11,22}(s)(k_j)n_{31,42}(s)(k_j)}[M_4(x,t,k_j)]_2e^{2\theta(k_j)}, \vspace{0.2in}\\
    & \quad {\rm for}\quad n_3+1\leq j\leq N,\quad k\in D_4, \quad l=3,4
   \end{array}  \ene
where the overdot denotes the derivative with resect to the parameter $k$ and $\theta=\theta(k)=-i(kx+2k^2t)$.}

\vspace{0.1in}

\noindent {\bf Proof.}\, It follows from Eqs.~(\ref{mns}) and (\ref{sn}) that we find the four columns of $M_1(x,t,k)$ as
\bes \label{m1}\bee
\label{m1a}  &[M_1]_1=[\mu_2]_1 \dfrac{m_{22}(s)}{n_{33,44}(s)}+[\mu_2]_2\dfrac{m_{12}(s)}{n_{33,44}(s)}, \qquad\qquad\quad\quad\,\vspace{0.1in}\\
\label{m1b} &[M_1]_2=[\mu_2]_1 \dfrac{m_{21}(s)}{n_{33,44}(s)}+[\mu_2]_2\dfrac{m_{11}(s)}{n_{33,44}(s)},\qquad\qquad\quad\quad\, \vspace{0.1in}\\
\label{m1c} &[M_1]_3=[\mu_2]_1 s_{13}e^{2\theta} +[\mu_2]_2s_{23}e^{2\theta}+[\mu_2]_3s_{33}+[\mu_2]_4s_{43}, \vspace{0.1in}\\
\label{m1d} &[M_1]_4=[\mu_2]_1 s_{14}e^{2\theta} +[\mu_2]_2s_{24}e^{2\theta}+[\mu_2]_3s_{34}+[\mu_2]_4s_{44},
  \ene \ees

For the case that $k_j\in D_1$ is a simple zero of $n_{33,44}(s)(k)$, it follows from Eqs.~(\ref{m1c}) and (\ref{m1d}) that we obtain $[\mu_2]_1$ and $[\mu_2]_2$ and then substitute them into Eqs.~(\ref{m1a}) and (\ref{m1b}) to yield
\bes\bee \nonumber
& [M_1]_1=\dfrac{m_{22}(s)s_{24}-m_{12}(s)s_{14}}{n_{33,44}(s)n_{13,24}(s)}[M_1]_3e^{-2\theta}
          +\dfrac{m_{12}(s)s_{13}-m_{22}(s)s_{23}}{n_{33,44}(s)n_{13,24}(s)}[M_1]_4e^{-2\theta} \vspace{0.1in}\\
  &  +\dfrac{m_{42}(s)[\mu_2]_3+m_{32}(s)[\mu_2]_4}{n_{13,24}(s)}e^{-2\theta},\vspace{0.1in}\\
&\nonumber
[M_1]_2=\dfrac{m_{21}(s)s_{24}-m_{11}(s)s_{14}}{n_{33,44}(s)n_{13,24}(s)}[M_1]_3e^{-2\theta}
          +\dfrac{m_{11}(s)s_{13}-m_{21}(s)s_{23}}{n_{33,44}(s)n_{13,24}(s)}[M_1]_4e^{-2\theta} \vspace{0.1in}\\
  &  +\dfrac{m_{41}(s)[\mu_2]_3+m_{31}(s)[\mu_2]_4}{n_{13,24}(s)}e^{-2\theta},
\ene\ees
whose residues at $k=k_j,\, k_j\in D_1$ yield Eq.~(\ref{rm1a}).

Similarly, we can show  Eq.~(\ref{rm2a}) for $k_j\in D_2$,  Eq.~(\ref{rm3a}) for $k_j\in D_3$, and Eq.~(\ref{rm4a}) for $k_j\in D_4$ by studying  Eqs.~(\ref{mns}) and (\ref{sn}) for $n=2,3,4$.  $\square$

\section{The $4\times 4$ matrix Riemann-Hilbert problem}

\quad By using the district contours $\gamma_j\, (j=1,2,3,4)$, the integral solutions of the revised Lax pair (\ref{mulax}), and $S_n$ due to $\{S(k), s(k)\}$, we have defined the sectionally analytic function $M_n(x,t,k),\, n=1,2,3,4$, which solves a
$4\times 4$ matrix Riemann-Hilbert (RH) problem. This RH problem can be formulated on basis of the initial conditions of the Schwartz class $q_j(x, t=0)$ and Dirichlet-Neumann boundary data $q_j(x=0,t)$ and $q_{jx}(x=0,t),\, j=1,0,-1$. Thus the solution of Eq.~(\ref{pnls}) for all values of $x,t$ can be refound by solving the RH problem.

\vspace{0.1in}
\noindent {\bf Theorem 3.1.}\, {\it Suppose that $(q_1(x,t),\, q_0(x,t),\, q_{-1}(x,t))$ is a solution of Eq.~(\ref{pnls}) in the domain  $\Omega=\{(x,t)\,| \, 0<x<\infty,\, t\in [0, T]\}$ with sufficient smoothness and decay as $x\to\infty$. Then it can be reconstructed from the initial data defined by $q_j(x, t=0)=q_{0j}(x), j=1,0,-1$ and Dirichlet and Neumann boundary values defined by
$q_j(x=0, t)=u_{0j}(t)$ and $q_{jx}(x=0, t)=u_{1j}(t),\, j=1,0, -1$.

We use the initial and boundary data to define the jump matrices $J_{mn}(x, t, k),\, n, m = 1,..., 4$, by Eq.~(\ref{jump}) as well as the
spectral functions $S(k), \, s(k)$  given by Eq. (\ref{sss}). Assume that the possible zeros $\{k_j\}^N_1$ of the functions $n_{33,44}(s)(k),\,  \mathcal{N}([S]_1[S]_2[s]_3[s]_4)(k),\, \mathcal{N}([s]_1[s]_2[S]_3[S]_4)(k)$ and $n_{11,22}(s)(k)$ are as in Assumption 2.4. Then the solution $(q_1(x,t),\, q_0(x,t), \, q_{-1}(x,t))$ of Eq.~(\ref{pnls}) is given by $M(x,t,k)$ in the form
\bee\label{solu}
\left\{\begin{array}{l}
q_1(x,t)=\d 2i\lim_{k\to \infty}(kM(x,t,k))_{13}, \vspace{0.1in}\\
 q_0(x,t)=\d 2i\lim_{k\to \infty}(kM(x,t,k))_{14}=2i\beta\lim_{k\to \infty}(kM(x,t,k))_{23},\vspace{0.1in}\\
q_{-1}(x,t)=\d 2i\lim_{k\to \infty}(kM(x,t,k))_{24},
\end{array}\right.
\ene
where $M(x,t,k)$ satisfies the following $4\times 4$ matrix Riemann-Hilbert problem:

\begin{itemize}
\item {} $M(x,t,k)$ is sectionally meromorphic on the Riemann $k$-sphere with jumps
 across the contours $\bar{D}_n\cup \bar{D}_m$, $(n, m = 1,2,3,4)$ (see Fig.~\ref{kplane}a).

\item {} Across the contours $\bar{D}_n\cup \bar{D}_m\, (n, m = 1,2,3,4)$, $M(x, t, k)$ satisfies the
 jump condition (\ref{jumpc}).

\item {} The residue conditions of $M(x,t,k)$ are satisfied in Proposition 2.4.

\item {} $M(x, t, k) = \d\mathbb{I}+O(1/k)$ as $k\to\infty$.

\end{itemize}
}

\noindent {\bf Proof.}\, System~(\ref{solu}) can be deduced from the large $k$ asymptotics of the eigenfunctions.
We can follow the similar one in Refs.~\cite{f3, nls3}  to show the rest proof of the Theorem. $\square$

\section{Nonlinearizable boundary conditions}

\quad The main difficulty of the initial-boundary value problems is to find the boundary values for a well-posed problem.
 All boundary conditions are required for the definition of $S(k)$, and hence for the formulate the $4\times 4$ matrix
 RH problem. Our main conclusion exhibits the unknown boundary condition on basis of the prescribed boundary condition and the initial condition
 in terms of the solution of a system of nonlinear integral equations.

\subsection*{(a) \, The time evolution of the global relation }

\quad By evaluating Eq.~(\ref{mur}) at $(x,t)=(0, t)$ and considering the global relation (\ref{srr2}), we have
\bee\label{gr}
 c(t,k)=\mu_2(0,t,k)e^{-2ik^2t\hat{\sigma}_4}s(k),\quad 0<t<T,\quad  k\in (C^-, C^-, C^+, C^+),
 \ene
which can be written as
\bes\bee
& \label{c12o}
\d [c(t,k)]_l=\sum_{j=1}^2[\mu_2(0,t,k)]_js_{jl}(k)+\sum_{j=3}^4[\mu_2(0,t,k)]_js_{jl}(k)e^{-4ik^2t},\quad l=1,2, \vspace{0.1in} \\
& \label{c34o}
\d [c(t,k)]_l=\sum_{j=1}^2[\mu_2(0,t,k)]_js_{jl}(k)e^{4ik^2t}+\sum_{j=3}^4[\mu_2(0,t,k)]_js_{jl}(k), \quad l=3,4,
    \ene\ees
Thus, the column vectors $[c(t,k)]_l,\, l=1,2$ are analytic and bounded in $C^-$ away from the possible zeros of $n_{11,22}(s)(k)$ and of order $O(1/k)$ as $k\to \infty$, and the column vectors $[c(t,k)]_l,\, l=3,4$ are analytic and bounded in $C^+$ away from the possible zeros of $n_{33,44}(s)(k)$ and of order $O(1/k)$ as $k\to \infty$.

\subsection*{(b) \,  Asymptotic behaviors of eigenfunctions}

\quad It follows from Eq.~(\ref{mulax}) that we have the asymptotics of eigenfunctions $\{\mu_j\}_1^3$ as $k\to\infty$
\bee\label{mua}
\begin{array}{rl}
\mu_j(x,t,k)=&\!\!\d\mathbb{I}+\sum_{s=1}^2
\frac{1}{k^s}\left(\begin{array}{cccc} \mu_{j,11}^{(s)} & \mu_{j,12}^{(s)} & \mu_{j,13}^{(s)} & \mu_{j,14}^{(s)} \vspace{0.1in} \\
                                      \mu_{j,21}^{(s)} & \mu_{j,22}^{(s)} & \mu_{j,23}^{(s)} & \mu_{j,24}^{(s)} \vspace{0.1in}\\
                                      \mu_{j,31}^{(s)} & \mu_{j,32}^{(s)} & \mu_{j,33}^{(s)} & \mu_{j,34}^{(s)} \vspace{0.1in}\\
                                      \mu_{j,41}^{(s)} & \mu_{j,42}^{(s)} & \mu_{j,43}^{(s)} & \mu_{j,44}^{(s)}
  \end{array}\right) +O(\frac{1}{k^3}) \vspace{0.2in}\\
 =&\!\! \d\mathbb{I}
 +\frac{1}{k}\left(\!\!\!\!\begin{array}{cccc} \int_{(x_j, t_j)}^{(x,t)}\Delta_{11}^{(1)} & \int_{(x_j, t_j)}^{(x,t)}\Delta_{12}^{(1)}
  &\!\!\!\!\! \d-\frac{i}{2}q_1 & \d-\frac{i}{2}q_0 \vspace{0.1in}\\
                                        \int_{(x_j, t_j)}^{(x,t)}\Delta_{21}^{(1)} & \int_{(x_j, t_j)}^{(x,t)}\Delta_{22}^{(1)} &
                                        \d-\frac{i\beta}{2}q_0 & \d-\frac{i}{2}q_{-1} \vspace{0.1in}\\
                                     \d \frac{i\alpha}{2}\bar{q}_1 & \d\frac{i\alpha\beta}{2}\bar{q}_0 &
                                     \int_{(x_j, t_j)}^{(x,t)}\Delta_{33}^{(1)} & \int_{(x_j, t_j)}^{(x,t)}\Delta_{34}^{(1)} \vspace{0.1in}\\
                                     \d \frac{i\alpha}{2}\bar{q}_0 & \d\frac{i\alpha}{2}\bar{q}_{-1} & \int_{(x_j, t_j)}^{(x,t)}\Delta_{43}^{(1)} & \int_{(x_j, t_j)}^{(x,t)}\Delta_{44}^{(1)}
  \end{array}\!\!\!\!\right) \vspace{0.1in}\\
   &\!\!\!\d +\frac{1}{k^2}\left(\begin{array}{cccc} \int_{(x_j, t_j)}^{(x,t)}\Delta_{11}^{(2)} & \!\!\int_{(x_j, t_j)}^{(x,t)}\Delta_{12}^{(2)}
                                            & \mu_{j,13}^{(2)} & \mu_{j,14}^{(2)} \vspace{0.1in}\\
                                      \int_{(x_j, t_j)}^{(x,t)}\Delta_{21}^{(2)} & \int_{(x_j, t_j)}^{(x,t)}\Delta_{22}^{(2)}
                                       & \mu_{j,23}^{(2)} & \mu_{j,24}^{(2)} \vspace{0.1in}\\
                                      \mu_{j,31}^{(2)} & \mu_{j,32}^{(2)} &
                                       \int_{(x_j, t_j)}^{(x,t)}\Delta_{33}^{(2)} & \int_{(x_j, t_j)}^{(x,t)}\Delta_{34}^{(2)} \vspace{0.1in}\\
                                      \mu_{j,41}^{(2)} & \mu_{j,42}^{(2)} &
                                      \int_{(x_j, t_j)}^{(x,t)}\Delta_{43}^{(2)} & \int_{(x_j, t_j)}^{(x,t)}\Delta_{44}^{(2)}
  \end{array}\!\!\!\right)+O(\frac{1}{k^3}),
  \end{array}
\ene
where we have introduced the following functions
\bee\no
\left\{\!\!\!\begin{array}{rl}
\Delta_{11}^{(1)}=&\!\!-\Delta_{33}^{(1)}
=\d\frac{i\alpha}{2}(|q_1|^2+|q_0|^2)dx+\frac{\alpha}{2}\sum_{j=0,1}(q_j\bar{q}_{jx}-q_{jx}\bar{q}_j)dt, \vspace{0.1in} \\
\Delta_{22}^{(1)}=&\!\!-\Delta_{44}^{(1)}=\d\frac{i\alpha}{2}(|q_{-1}|^2+|q_0|^2)dx+\frac{\alpha}{2}\sum_{j=-1,0}(q_j\bar{q}_{jx}-q_{jx}\bar{q}_j)dt, \vspace{0.1in} \\
\Delta_{12}^{(1)}=&\!\!\!\! -\bar{\Delta}_{21}^{(1)}=-\Delta_{34}^{(1)}=\bar{\Delta}_{43}^{(1)}=\frac{i\alpha}{2}(\beta q_1\bar{q}_0+q_0\bar{q}_{-1})dx+\frac{\alpha}{2}
 (\beta q_1\bar{q}_{0x}-\beta q_{1x}\bar{q}_0+q_0\bar{q}_{-1x}-q_{0x}\bar{q}_{-1})dt,
\end{array}\right.
\ene
\bee\no
\left\{\begin{array}{rl}
\mu_{j,13}^{(2)}=&\!\!\d\frac{1}{4}q_{1x}+\frac{1}{2i}\left(q_1\mu_{j,33}^{(1)}+q_0\mu_{j,43}^{(1)}\right) =\frac{1}{4}q_{1x}+\frac{1}{2i}\left[q_1\int_{(x_j,t_j)}^{(x,t)}\Delta_{33}^{(1)}+q_0\int_{(x_j,t_j)}^{(x,t)}\Delta_{43}^{(1)}\right], \vspace{0.1in} \\
\mu_{j,14}^{(2)}=&\!\!\d\frac{1}{4}q_{0x}+\frac{1}{2i}\left(q_1\mu_{j,34}^{(1)}+q_0\mu_{j,44}^{(1)}\right) =\frac{1}{4}q_{0x}+\frac{1}{2i}\left[q_1\int_{(x_j,t_j)}^{(x,t)}\Delta_{34}^{(1)}+q_0\int_{(x_j,t_j)}^{(x,t)}\Delta_{44}^{(1)}\right], \vspace{0.1in} \\
\mu_{j,23}^{(2)}=&\!\!\d\frac{\beta}{4}q_{0x}+\frac{1}{2i}\left(\beta q_0\mu_{j,33}^{(1)}+q_{-1}\mu_{j,43}^{(1)}\right)=\frac{\beta}{4}q_{0x}+\frac{1}{2i}\left[\beta q_0\int_{(x_j,t_j)}^{(x,t)}\Delta_{33}^{(1)}+q_{-1}\int_{(x_j,t_j)}^{(x,t)}\Delta_{43}^{(1)}\right], \vspace{0.1in} \\
\mu_{j,24}^{(2)}=&\!\!\d\frac{1}{4}q_{-1x}+\frac{1}{2i}\left(\beta q_0\mu_{j,34}^{(1)}+q_{-1}\mu_{j,44}^{(1)}\right)=\frac{1}{4}q_{-1x}+\frac{1}{2i}\left[\beta q_0\int_{(x_j,t_j)}^{(x,t)}\Delta_{34}^{(1)}+q_{-1}\int_{(x_j,t_j)}^{(x,t)}\Delta_{44}^{(1)}\right],
 \end{array}\right.
\ene
\bee\no
\left\{\begin{array}{rl}
\mu_{j,31}^{(2)}=&\!\!\d\frac{\alpha}{4}\bar{q}_{1x}+\frac{i\alpha }{2}\left(\bar{q}_1\mu_{j,11}^{(1)}+\beta\bar{q}_0\mu_{j,21}^{(1)}\right) =\frac{\alpha}{4}\bar{q}_{1x}+\frac{i\alpha}{2}\left[\bar{q}_1\int_{(x_j,t_j)}^{(x,t)}\Delta_{11}^{(1)}+\beta\bar{q}_0
    \int_{(x_j,t_j)}^{(x,t)}\Delta_{21}^{(1)}\right], \vspace{0.1in} \\
\mu_{j,32}^{(2)}=&\!\!\d\frac{\alpha\beta}{4}\bar{q}_{0x}+\frac{i\alpha }{2}\left(\bar{q}_1\mu_{j,12}^{(1)}+\beta\bar{q}_0\mu_{j,22}^{(1)}\right) =\frac{\alpha\beta}{4}\bar{q}_{0x}+\frac{i\alpha}{2}\left[\bar{q}_1\int_{(x_j,t_j)}^{(x,t)}\Delta_{12}^{(1)}+\beta\bar{q}_0
    \int_{(x_j,t_j)}^{(x,t)}\Delta_{22}^{(1)}\right], \vspace{0.1in} \\
\mu_{j,41}^{(2)}=&\!\!\d\frac{\alpha}{4}\bar{q}_{0x}+\frac{i\alpha }{2}\left(\bar{q}_0\mu_{j,11}^{(1)}+\bar{q}_{-1}\mu_{j,21}^{(1)}\right) =\frac{\alpha}{4}\bar{q}_{0x}+\frac{i\alpha}{2}\left[\bar{q}_0\int_{(x_j,t_j)}^{(x,t)}\Delta_{11}^{(1)}+\beta\bar{q}_{-1}
    \int_{(x_j,t_j)}^{(x,t)}\Delta_{21}^{(1)}\right], \vspace{0.1in} \\
\mu_{j,42}^{(2)}=&\!\!\d\frac{\alpha}{4}\bar{q}_{-1x}+\frac{i\alpha }{2}\left(\bar{q}_0\mu_{j,12}^{(1)}+\bar{q}_{-1}\mu_{j,22}^{(1)}\right) =\frac{\alpha}{4}\bar{q}_{-1x}+\frac{i\alpha}{2}\left[\bar{q}_0\int_{(x_j,t_j)}^{(x,t)}\Delta_{12}^{(1)}+\beta\bar{q}_{-1}
    \int_{(x_j,t_j)}^{(x,t)}\Delta_{22}^{(1)}\right],
\end{array}\right.
\ene
\bee\begin{array}{rl}
\Delta_{11}^{(2)}=&\!\! \d\left\{\frac{\alpha}{4}(q_1\bar{q}_{1x}+q_0\bar{q}_{0x})
   +\frac{i\alpha}{2}\left[(|q_1|^2+|q_0|^2)\mu_{j,11}^{(1)}+(\beta q_1\bar{q}_0+q_0\bar{q}_{-1})\mu_{j,21}^{(1)}\right]\right\}dx \vspace{0.1in}\\
  &\!\!\d +\left\{\frac{ \alpha}{4}(q_1\bar{q}_{1t}+q_0\bar{q}_{0t})+\frac{i\alpha}{4}(q_{1x}\bar{q}_{1x}+q_{0x}\bar{q}_{0x})\right.  -\frac{i}{4}[(|q_1|^2+|q_0|^2)^2  \vspace{0.1in}\\
  &\!\!\d + (\beta q_1\bar{q}_0+q_0\bar{q}_{-1})(\beta q_0\bar{q}_{1}+q_{-1}\bar{q}_{0})] +\frac{\alpha}{2}(q_1\bar{q}_{1x}-q_{1x}\bar{q}_1+q_0\bar{q}_{0x}-q_{0x}\bar{q}_0)\mu_{j,11}^{(1)} \vspace{0.1in} \\
  &\!\!\d \left.+\frac{\alpha}{2}(\beta q_1\bar{q}_{0x}-\beta q_{1x}\bar{q}_0+q_0\bar{q}_{-1x}-q_{0x}\bar{q}_{-1})\mu_{j,21}^{(1)}\right\} dt,
  \end{array} \\
\begin{array}{rl}
\Delta_{12}^{(2)}=&\!\!\d \left\{\frac{\alpha}{4}(\beta q_1\bar{q}_{0x}+q_0\bar{q}_{-1x})
  +\frac{i\alpha}{2}\left[(|q_1|^2+|q_0|^2)\mu_{j,12}^{(1)}+(\beta q_1\bar{q}_0+q_0\bar{q}_{-1})\mu_{j,22}^{(1)}\right]\right\}dx \vspace{0.1in}\\
 &\!\!\d +\left\{\frac{ \alpha}{4}(\beta q_1\bar{q}_{0t}+q_0\bar{q}_{-1t})-\frac{i}{4}(\beta q_1\bar{q}_0+q_0\bar{q}_{-1})(|q_1|^2+2|q_0|^2+|q_{-1}|^2) \right. \vspace{0.1in}\\
  &\d +\frac{i\alpha}{4}(\beta q_{1x}\bar{q}_{0x}+q_{0x}\bar{q}_{-1x})+\frac{\alpha}{2}(q_1\bar{q}_{1x}-q_{1x}\bar{q}_1+q_0\bar{q}_{0x}-q_{0x}\bar{q}_0)\mu_{j,12}^{(1)} \vspace{0.1in} \\
  &\d \left.+\frac{\alpha}{2}(\beta q_1\bar{q}_{0x}-\beta q_{1x}\bar{q}_0+q_0\bar{q}_{-1x}-q_{0x}\bar{q}_{-1})\mu_{j,22}^{(1)}\right\} dt,
  \end{array} \\
 \begin{array}{rl}
 \Delta_{21}^{(2)}=&\!\!\d\left\{\frac{\alpha}{4}(\beta q_0\bar{q}_{1x}+q_{-1}\bar{q}_{0x})
   +\frac{i\alpha}{2}\left[(\beta q_0\bar{q}_{1}+q_{-1}\bar{q}_{0})\mu_{j,11}^{(1)}+(|q_{-1}|^2+|q_0|^2)\mu_{j,21}^{(1)}\right]\right\}dx \\
   &\!\!\d +\left\{\frac{ \alpha}{4}(\beta q_0\bar{q}_{1t}+q_{-1}\bar{q}_{0t})
    -\frac{i}{4}(\beta q_0\bar{q}_1+q_{-1}\bar{q}_0)(|q_1|^2+2|q_0|^2+|q_{-1}|^2)\right. \vspace{0.1in}\\
  &\!\!\d +\frac{i\alpha}{4}(\beta q_{0x}\bar{q}_{1x}+q_{-1x}\bar{q}_{0x})
  +\frac{\alpha}{2}(q_{-1}\bar{q}_{-1x}-q_{-1x}\bar{q}_{-1}+q_0\bar{q}_{0x}-q_{0x}\bar{q}_0)\mu_{j,21}^{(1)} \vspace{0.1in} \\
  &\!\!\d \left.+\frac{\alpha}{2}(\beta q_0\bar{q}_{1x}-\beta q_{0x}\bar{q}_1+q_{-1}\bar{q}_{0x}-q_{-1x}\bar{q}_0)\mu_{j,11}^{(1)}\right\} dt,
  \end{array} \\
\begin{array}{rl}
 \Delta_{22}^{(2)}=&\!\!\d \left\{\frac{\alpha}{4}(q_{-1}\bar{q}_{-1x}+q_0\bar{q}_{0x})+\frac{i\alpha}{2}\left[(\beta q_0\bar{q}_{1}+q_{-1}\bar{q}_{0})\mu_{j,12}^{(1)}+(|q_{-1}|^2+|q_0|^2)\mu_{j,22}^{(1)}\right]\right\}dx \\
  &\!\!\d +\left\{\frac{ \alpha}{4}(q_{-1}\bar{q}_{-1t}+q_0\bar{q}_{0t})
   +\frac{i\alpha}{4}(q_{-1x}\bar{q}_{-1x}+q_{0x}\bar{q}_{0x})-\frac{i}{4}[|q_0|^2+|q_{-1}|^2)^2\right.  \vspace{0.1in}\\
  &\d +(\beta q_0\bar{q}_1+q_{-1}\bar{q}_0)(\beta q_1\bar{q}_0+q_0\bar{q}_{-1})] +\frac{\alpha}{2}(q_1\bar{q}_{-1x}-q_{-1x}\bar{q}_{-1}+q_0\bar{q}_{0x}-q_{0x}\bar{q}_0)\mu_{j,22}^{(1)} \vspace{0.1in} \\
  &\!\!\d \left.+\frac{\alpha}{2}(\beta q_0\bar{q}_{1x}-\beta q_{0x}\bar{q}_1+q_{-1}\bar{q}_{0x}-q_{-1x}\bar{q}_0)\mu_{j,12}^{(1)}\right\} dt,
  \end{array}
\ene

\bee\begin{array}{rl}
\Delta_{33}^{(2)}=&\!\! \d\left\{\frac{\alpha}{4}(q_{1x}\bar{q}_{1}+q_{0x}\bar{q}_{0})
   -\frac{i\alpha}{2}\left[(|q_1|^2+|q_0|^2)\mu_{j,33}^{(1)}+(\beta q_{-1}\bar{q}_0+q_0\bar{q}_{1})\mu_{j,43}^{(1)}\right]\right\}dx \\
    &\!\!\d +\left\{\frac{ \alpha}{4}(q_{1t}\bar{q}_{1}+q_{0t}\bar{q}_{0})-\frac{i\alpha}{4}(q_{1x}\bar{q}_{1x}+q_{0x}\bar{q}_{0x}) +\frac{i}{4}[(|q_1|^2+|q_0|^2)^2  \right.  \vspace{0.1in}\\
  & \d + (\beta q_{-1}\bar{q}_0+q_0\bar{q}_{1})(\beta q_0\bar{q}_{-1}+q_{1}\bar{q}_{0})] +\frac{\alpha}{2}(q_{1x}\bar{q}_{1}-q_{1}\bar{q}_{1x}+q_{0x}\bar{q}_{x}-q_{0}\bar{q}_{0x})\mu_{j,33}^{(1)} \vspace{0.1in} \\
  &\!\!\d \left.+\frac{\alpha}{2}(\beta q_{-1x}\bar{q}_{0}-\beta q_{-1}\bar{q}_{0x}+q_{0x}\bar{q}_{1}-q_{0}\bar{q}_{1x})\mu_{j,43}^{(1)}\right\} dt,
  \end{array} \\
\begin{array}{rl}
\Delta_{34}^{(2)}=&\!\!\d \left\{\frac{\alpha}{4}(\beta q_{-1x}\bar{q}_{0}+q_0\bar{q}_{1x})
  -\frac{i\alpha}{2}\left[(|q_1|^2+|q_0|^2)\mu_{j,34}^{(1)}+(\beta q_{-1}\bar{q}_0+q_0\bar{q}_{1})\mu_{j,44}^{(1)}\right]\right\}dx \\
    &\!\!\d +\left\{\frac{ \alpha}{4}(\beta q_{-1t}\bar{q}_0+q_{0t}\bar{q}_1)
    +\frac{i}{4}(\beta q_{-1}\bar{q}_0+q_0\bar{q}_{1})(|q_1|^2+2|q_0|^2+|q_{-1}|^2)  \right.  \vspace{0.1in}\\
  &\!\!\d -\frac{i\alpha}{4}(q_{-1x}\bar{q}_{0x}+q_{0x}\bar{q}_{1x})
  +\frac{\alpha}{2}(q_{1x}\bar{q}_{1}-q_{1}\bar{q}_{1x}+q_{0x}\bar{q}_{x}-q_{0}\bar{q}_{0x})\mu_{j,34}^{(1)} \vspace{0.1in} \\
  &\!\!\d \left.+\frac{\alpha}{2}(\beta q_{-1x}\bar{q}_{0}-\beta q_{-1}\bar{q}_{0x}+q_{0x}\bar{q}_{1}-q_{0}\bar{q}_{1x})\mu_{j,44}^{(1)}\right\} dt,
  \end{array} \\
\begin{array}{rl}
\Delta_{43}^{(2)}=&\!\!\d\left\{\frac{\alpha}{4}(\beta q_{0x}\bar{q}_{1}+q_{1x}\bar{q}_{0})
   -\frac{i\alpha}{2}\left[(\beta q_0\bar{q}_{-1}+q_{1}\bar{q}_{0})\mu_{j,33}^{(1)}+(|q_{-1}|^2+|q_0|^2)\mu_{j,43}^{(1)}\right]\right\}dx \\
   &\!\!\d +\left\{\frac{ \alpha}{4}(\beta q_{0t}\bar{q}_{-1}+q_{1t}\bar{q}_0)
     +\frac{i}{4}(\beta q_0\bar{q}_{-1}+q_1\bar{q}_0)(|q_1|^2+2|q_0|^2+|q_{-1}|^2) \right.  \vspace{0.1in}\\
  &\!\!\d -\frac{i\alpha}{4}(\beta q_{0x}\bar{q}_{-1x}+q_{1x}\bar{q}_{0x})+\frac{\alpha}{2}(q_{-1x}\bar{q}_{-1}-q_{-1}\bar{q}_{-1x}+q_{0x}\bar{q}_{x}-q_{0}\bar{q}_{0x})\mu_{j,43}^{(1)} \vspace{0.1in} \\
  &\!\!\d \left.+\frac{\alpha}{2}(\beta q_{0x}\bar{q}_{-1}-\beta q_0\bar{q}_{-1x}+q_{1x}\bar{q}_{0}-q_{1}\bar{q}_{0x})\mu_{j,33}^{(1)}\right\} dt,
  \end{array} \\
\begin{array}{rl}
\Delta_{44}^{(2)}=&\!\!\d \left\{\frac{\alpha}{4}(q_{-1x}\bar{q}_{-1}+q_{0x}\bar{q}_{0})
   -\frac{i\alpha}{2}\left[(\beta q_0\bar{q}_{-1}+q_{1}\bar{q}_{0})\mu_{j,34}^{(1)}+(|q_{-1}|^2+|q_0|^2)\mu_{j,44}^{(1)}\right]\right\}dx \\
     &\!\!\d +\left\{\frac{ \alpha}{4}(q_{-1t}\bar{q}_{-1}+q_{0t}\bar{q}_0)
    +\frac{i}{4}[(\beta q_{-1}\bar{q}_0+q_0\bar{q}_1)(\beta q_0\bar{q}_{-1}+q_1\bar{q}_0)+(|q_0|^2+|q_{-1}|^2)^2]  \right.  \vspace{0.1in}\\
  &\!\!\d -\frac{i\alpha}{4}(q_{-1x}\bar{q}_{-1x}+q_{0x}\bar{q}_{0x})+\frac{\alpha}{2}(q_{-1x}\bar{q}_{-1}-q_{-1}\bar{q}_{-1x}+q_{0x}\bar{q}_{x}-q_{0}\bar{q}_{0x})\mu_{j,44}^{(1)} \vspace{0.1in} \\
  &\!\!\d \left.+\frac{\alpha}{2}(\beta q_{0x}\bar{q}_{-1}-\beta q_0\bar{q}_{-1x}+q_{1x}\bar{q}_{0}-q_{1}\bar{q}_{0x})\mu_{j,34}^{(1)}\right\} dt,
  \end{array}
\ene
where the functions $\{\mu^{(i)}_{jl}=\mu^{(i)}_{jl}(x,t)\}_1^3,\, i=1, 2$ are independent of $k$.

We define the matrix-valued function $\Psi(t,k)=(\Psi_{ij}(t,k))_{4\times 4}$ as
\bee\label{mu2asy}
\begin{array}{rl}
\mu_2(0, t,k)=\Psi(t, k)=&\d \mathbb{I}+\sum_{s=1}^2\frac{1}{k^s}\left(\begin{array}{cccc}
 \Psi_{11}^{(s)}(t) & \Psi_{12}^{(s)}(t) & \Psi_{13}^{(s)}(t) & \Psi_{14}^{(s)}(t) \vspace{0.1in} \\
 \Psi_{21}^{(s)}(t) & \Psi_{22}^{(s)}(t) & \Psi_{23}^{(s)}(t) & \Psi_{24}^{(s)}(t) \vspace{0.1in} \\
 \Psi_{31}^{(s)}(t) & \Psi_{32}^{(s)}(t) & \Psi_{33}^{(s)}(t) & \Psi_{34}^{(s)}(t) \vspace{0.1in} \\
 \Psi_{41}^{(s)}(t) & \Psi_{42}^{(s)}(t) & \Psi_{43}^{(s)}(t) & \Psi_{44}^{(s)}(t)
 \end{array}
\right) +O(\frac{1}{k^3}),
\end{array}
\ene

By using the asymptotic of Eq.~(\ref{mua}) and the boundary data at $x=0$, we find
\bee\left\{\begin{array}{l}
\Psi_{13}^{(1)}(t)=-\d\frac{i}{2}u_{01}(t), \quad \Psi_{14}^{(1)}(t)=\beta\Psi_{23}^{(1)}(t)=-\frac{i}{2}u_{00}(t),\quad
\Psi_{24}^{(1)}(t)=-\d\frac{i}{2}u_{0-1}(t),
 \vspace{0.1in} \\
\Psi_{13}^{(2)}(t)=\d\frac{1}{4}u_{11}(t)-\frac{i}{2}\left[u_{01}(t)\Psi_{33}^{(1)}+u_{00}(t)\Psi_{43}^{(1)}\right], \vspace{0.1in} \\
\Psi_{14}^{(2)}(t)=\d\frac{1}{4}u_{10}(t)-\frac{i}{2}\left[u_{01}(t)\Psi_{34}^{(1)}+u_{00}(t)\Psi_{44}^{(1)}\right], \vspace{0.1in} \\
\Psi_{23}^{(2)}(t)=\d\frac{\beta}{4}u_{10}(t)-\frac{i}{2}\left[\beta u_{00}(t)\Psi_{33}^{(1)}+u_{0-1}(t)\Psi_{43}^{(1)}\right], \vspace{0.1in} \\
\Psi_{24}^{(2)}(t)=\d\frac{1}{4}u_{1-1}(t)-\frac{i}{2}\left[\beta u_{00}(t)\Psi_{34}^{(1)}+u_{0-1}(t)\Psi_{44}^{(1)}\right], \vspace{0.1in} \\
\Psi_{33}^{(1)}(t)=\d\frac{\alpha}{2}\int^t_0\sum_{j=0,1}\left[\bar{u}_{0j}(t)u_{1j}(t)-u_{0j}(t)\bar{u}_{1j}(t)\right]dt, \vspace{0.1in} \\
\Psi_{44}^{(1)}(t)=\d\frac{\alpha}{2}\int^t_0\sum_{j=-1, 0}\left[\bar{u}_{0j}(t)u_{1j}(t)-u_{0j}(t)\bar{u}_{1j}(t)\right]dt, \vspace{0.1in} \\
\Psi_{34}^{(1)}(t)=\displaystyle\frac{\alpha}{2}\int^t_0\left[\beta u_{11}(t)\bar{u}_{00}(t)-\beta u_{01}(t)\bar{u}_{10}(t)+u_{10}(t)\bar{u}_{0-1}(t)-u_{00}(t)\bar{u}_{1-1}(t)\right]dt,\vspace{0.1in}  \\
\Psi_{43}^{(1)}(t)=\displaystyle\frac{\alpha}{2}\int^t_0\left[\beta u_{10}(t)\bar{u}_{01}(t)-\beta u_{00}(t)\bar{u}_{11}(t)+u_{1-1}(t)\bar{u}_{00}(t)-u_{0-1}(t)\bar{u}_{10}(t)\right]dt,\vspace{0.1in} \\
\end{array}\right.
\ene

Thus we have the the Dirichlet-Neumann boundary data at $x=0$:
\bee\label{ud}
\left\{\begin{array}{rl}
u_{01}(t)=&2i\Psi_{13}^{(1)}(t), \quad
u_{00}(t)=2i\Psi_{14}^{(1)}(t)=2i\beta\Psi_{23}^{(1)}(t),   \quad
u_{0-1}(t)= 2i\Psi_{24}^{(1)}(t),  \vspace{0.1in} \\
u_{11}(t)=&4\Psi_{13}^{(2)}(t)+2i[u_{01}(t)\Psi_{33}^{(1)}(t)+u_{00}(t)\Psi_{43}^{(1)}(t)] \vspace{0.1in} \\
u_{1-1}(t)=&4\Psi_{24}^{(2)}(t)+2i[\beta u_{00}(t)\Psi_{34}^{(1)}(t)+u_{0-1}(t)\Psi_{44}^{(1)}(t)], \vspace{0.1in} \\
u_{10}(t)=&4\Psi_{14}^{(2)}(t)+2i[u_{01}(t)\Psi_{34}^{(1)}(t)+u_{00}(t)\Psi_{44}^{(1)}(t)] \vspace{0.1in} \\
 =&4\beta\Psi_{23}^{(2)}(t)+2i\beta[\beta u_{00}(t)\Psi_{33}^{(1)}(t)+u_{0-1}(t)\Psi_{43}^{(1)}(t)],
\end{array}\right.
\ene

For the vanishing initial values, it follows from Eqs.~(\ref{c12o}) and (\ref{c34o}) that we have the following asymptotic of $c_{24}(t,k)$ and $c_{1j}(t,k), j=3,4$.

\vspace{0.1in}
\noindent {\bf Proposition 4.1.} {\it The global relation (\ref{gr}) implies that the large $k$ behavior of
 $c_{1j}(t,k),\, j=3,4$ and $c_{24}(t,k)$ is of the form }
\bes\bee
&\label{c13}
c_{13}(t,k)=\displaystyle\frac{\Psi_{13}^{(1)}}{k}+\frac{\Psi_{13}^{(2)}}{k^2}+O\left(\frac{1}{k^3}\right), \vspace{0.1in}\\
&\label{c14}
c_{14}(t,k)=\displaystyle\frac{\Psi_{14}^{(1)}}{k}+\frac{\Psi_{14}^{(2)}}{k^2}+O\left(\frac{1}{k^3}\right), \vspace{0.08in}\\
&\label{c24}
c_{24}(t,k)=\displaystyle\frac{\Psi_{24}^{(1)}}{k}+\frac{\Psi_{24}^{(2)}}{k^2}+O\left(\frac{1}{k^3}\right),
\ene
\ees

\vspace{0.1in}
\noindent {\bf Proof.}  The global relation (\ref{gr}) can be written as
\bes\bee
&  \label{ca}  c_{13}(t,k)=[\Psi_{11}(t,k)s_{13}+\Psi_{12}(t,k)s_{23}]e^{-4ik^2t}+\Psi_{13}(t,k)s_{33}+\Psi_{14}(t,k)s_{43}, \vspace{0.1in}\\
&  \label{cb}  c_{14}(t,k)=[\Psi_{11}(t,k)s_{14}+\Psi_{12}(t,k)s_{24}]e^{-4ik^2t}+\Psi_{13}(t,k)s_{34}+\Psi_{14}(t,k)s_{44}, \vspace{0.1in}\\
&  \label{cc}  c_{24}(t,k)=[\Psi_{21}(t,k)s_{14}+\Psi_{22}(t,k)s_{24}]e^{-4ik^2t}+\Psi_{23}(t,k)s_{34}+\Psi_{24}(t,k)s_{44},
\ene\ees

According to the asymptotics (\ref{mua}), we have
\bee\label{s13}
\left(\begin{array}{c} s_{13} \\ s_{23} \\ s_{33} \\ s_{43} \end{array}\right)
 =\left(\begin{array}{c} 0 \\ 0 \\ 1 \\ 0\end{array}\right)
 +\frac{1}{2ik}\left(\begin{array}{c} q_1(0,0) \vspace{0.1in}\\ \beta q_0(0,0,) \vspace{0.1in}\\
  2i\int_{(\infty, 0)}^{(0,0)}\Delta_{33}^{(1)}(0,0) \vspace{0.1in}\\ 2i\int_{(\infty, 0)}^{(0,0)}\Delta_{34}^{(1)}(0,0) \end{array}\right)
  +O(\frac{1}{k^2}),
\ene
and
\bee\label{s14}
\left(\begin{array}{c} s_{14} \\ s_{24} \\ s_{34} \\ s_{44} \end{array}\right)
 =\left(\begin{array}{c} 0 \\ 0 \\ 0 \\ 1 \end{array}\right)
 +\frac{1}{2ik}\left(\begin{array}{c} q_1(0,0) \vspace{0.1in}\\ \beta q_0(0,0,) \vspace{0.1in}\\
  2i\int_{(\infty, 0)}^{(0,0)}\Delta_{34}^{(1)}(0,0) \vspace{0.1in}\\ 2i\int_{(\infty, 0)}^{(0,0)}\Delta_{44}^{(1)}(0,0) \end{array}\right)
  +O(\frac{1}{k^2}),
\ene

Recalling the time-part of the Lax pair (\ref{mulax})
\bee\label{tlax}
 \mu_t+2ik^2[\sigma_4, \mu]=V(x,t,k)\mu,
 \ene
It follows from the first column of Eq.~(\ref{tlax}) with $\mu=\mu_2(0,t,k)=\Psi(t,k)$ that we have
\bee\label{psit11}
\left\{ \begin{array}{rl}
 \Psi_{11,t}(t,k)=& 2k(u_{01}\Psi_{31}+u_{00}\Psi_{41})+i(u_{11}\Psi_{31}+u_{10}\Psi_{41}) \vspace{0.08in}\\
  &-i\alpha[(|u_{01}|^2+|u_{00}|^2)\Psi_{11}+(\beta u_{01}\bar{u}_{02}+u_{00}\bar{u}_{0-1})\Psi_{21}], \vspace{0.08in}\\
 \Psi_{21,t}(t,k)=& 2k(\beta u_{00}\Psi_{31}+u_{0-1}\Psi_{41})+i(\beta u_{10}\Psi_{31}+u_{1-1}\Psi_{41}) \vspace{0.08in}\\
     & -i\alpha[(\beta u_{00}\bar{u}_{01}+u_{0-1}\bar{u}_{00})\Psi_{11}+(|u_{0-1}|^2+|u_{00}|^2)\Psi_{21}],\vspace{0.08in}\\
 \Psi_{31,t}(t,k)=&4ik^2\Psi_{31}+2\alpha k(\bar{u}_{01}\Psi_{11}+\beta \bar{u}_{00}\Psi_{21})
      -i\alpha(\bar{u}_{11}\Psi_{11}+\beta \bar{u}_{10}\Psi_{21}) \vspace{0.08in}\\
    & +i\alpha[(|u_{01}|^2+|u_{00}|^2)\Psi_{31}+(\beta u_{0-1}\bar{u}_{00}+u_{00}\bar{u}_{01})\Psi_{41}] \vspace{0.08in}\\
 \Psi_{41,t}(t,k)=&4ik^2\Psi_{41}+2\alpha k(\bar{u}_{00}\Psi_{11}+\bar{u}_{0-1}\Psi_{21})
   -i\alpha(\bar{u}_{10}\Psi_{11}+\bar{u}_{1-1}\Psi_{21})\vspace{0.08in}\\
& +i\alpha[(\beta u_{00}\bar{u}_{0-1}+u_{01}\bar{u}_{00})\Psi_{31}+(|u_{0-1}|^2+|u_{00}|^2)\Psi_{41}],
\end{array}\right.
\ene

The second column of Eq.~(\ref{tlax}) with $\mu=\mu_2(0,t,k)=\Psi(t,k)$ yields
\bee\label{psit12}
\left\{ \begin{array}{rl}
 \Psi_{12,t}(t,k)=& 2k(u_{01}\Psi_{32}+u_{00}\Psi_{42})+i(u_{11}\Psi_{32}+u_{10}\Psi_{42}) \vspace{0.08in}\\
  &-i\alpha[(|u_{01}|^2+|u_{00}|^2)\Psi_{12}+(\beta u_{01}\bar{u}_{02}+u_{00}\bar{u}_{0-1})\Psi_{22}], \vspace{0.08in}\\
 \Psi_{22,t}(t,k)=& 2k(\beta u_{00}\Psi_{32}+u_{0-1}\Psi_{42})+i(\beta u_{10}\Psi_{32}+u_{1-1}\Psi_{42}) \vspace{0.08in}\\
     & -i\alpha[(\beta u_{00}\bar{u}_{01}+u_{0-1}\bar{u}_{00})\Psi_{12}+(|u_{0-1}|^2+|u_{00}|^2)\Psi_{22}],\vspace{0.08in}\\
 \Psi_{32,t}(t,k)=&4ik^2\Psi_{32}+2\alpha k(\bar{u}_{01}\Psi_{12}+\beta \bar{u}_{00}\Psi_{22})
      -i\alpha(\bar{u}_{11}\Psi_{12}+\beta \bar{u}_{10}\Psi_{22}) \vspace{0.08in}\\
    & +i\alpha[(|u_{01}|^2+|u_{00}|^2)\Psi_{32}+(\beta u_{0-1}\bar{u}_{00}+u_{00}\bar{u}_{01})\Psi_{42}] \vspace{0.08in}\\
 \Psi_{42,t}(t,k)=&4ik^2\Psi_{42}+2\alpha k(\bar{u}_{00}\Psi_{12}+\bar{u}_{0-1}\Psi_{22})
   -i\alpha(\bar{u}_{10}\Psi_{12}+\bar{u}_{1-1}\Psi_{22})\vspace{0.08in}\\
& +i\alpha[(\beta u_{00}\bar{u}_{0-1}+u_{01}\bar{u}_{00})\Psi_{32}+(|u_{0-1}|^2+|u_{00}|^2)\Psi_{42}],
\end{array}\right.
\ene

The third column of Eq.~(\ref{tlax}) with $\mu=\mu_2(0,t,k)=\Psi(t,k)$ yields
\bee\label{psit13}
\left\{ \begin{array}{rl}
 \Psi_{13,t}(t,k)=&-4ik^2\Psi_{13}+2k(u_{01}\Psi_{33}+u_{00}\Psi_{43})+i(u_{11}\Psi_{33}+u_{10}\Psi_{43}) \vspace{0.08in}\\
  &-i\alpha[(|u_{01}|^2+|u_{00}|^2)\Psi_{13}+(\beta u_{01}\bar{u}_{02}+u_{00}\bar{u}_{0-1})\Psi_{23}], \vspace{0.08in}\\
 \Psi_{23,t}(t,k)=&-4ik^2\Psi_{23}+2k(\beta u_{00}\Psi_{33}+u_{0-1}\Psi_{43})+i(\beta u_{10}\Psi_{33}+u_{1-1}\Psi_{43}) \vspace{0.08in}\\
     & -i\alpha[(\beta u_{00}\bar{u}_{01}+u_{0-1}\bar{u}_{00})\Psi_{13}+(|u_{0-1}|^2+|u_{00}|^2)\Psi_{23}],\vspace{0.08in}\\
 \Psi_{33,t}(t,k)=&2\alpha k(\bar{u}_{01}\Psi_{13}+\beta \bar{u}_{00}\Psi_{23})
      -i\alpha(\bar{u}_{11}\Psi_{13}+\beta \bar{u}_{10}\Psi_{23}) \vspace{0.08in}\\
    & +i\alpha[(|u_{01}|^2+|u_{00}|^2)\Psi_{33}+(\beta u_{0-1}\bar{u}_{00}+u_{00}\bar{u}_{01})\Psi_{43}] \vspace{0.08in}\\
 \Psi_{43,t}(t,k)=&2\alpha k(\bar{u}_{00}\Psi_{13}+\bar{u}_{0-1}\Psi_{23})
   -i\alpha(\bar{u}_{10}\Psi_{13}+\bar{u}_{1-1}\Psi_{23})\vspace{0.08in}\\
& +i\alpha[(\beta u_{00}\bar{u}_{0-1}+u_{01}\bar{u}_{00})\Psi_{33}+(|u_{0-1}|^2+|u_{00}|^2)\Psi_{43}],
\end{array}\right.
\ene
and the fourth column of Eq.~(\ref{tlax}) with $\mu=\mu_2(0,t,k)=\Psi(t,k)$ yields
\bee\label{psit14}
\left\{ \begin{array}{rl}
 \Psi_{14,t}(t,k)=&-4ik^2\Psi_{14}+2k(u_{01}\Psi_{34}+u_{00}\Psi_{44})+i(u_{11}\Psi_{34}+u_{10}\Psi_{44}) \vspace{0.08in}\\
  &-i\alpha[(|u_{01}|^2+|u_{00}|^2)\Psi_{14}+(\beta u_{01}\bar{u}_{02}+u_{00}\bar{u}_{0-1})\Psi_{24}], \vspace{0.08in}\\
 \Psi_{24,t}(t,k)=&-4ik^2\Psi_{24}+2k(\beta u_{00}\Psi_{34}+u_{0-1}\Psi_{44})+i(\beta u_{10}\Psi_{34}+u_{1-1}\Psi_{44}) \vspace{0.08in}\\
     & -i\alpha[(\beta u_{00}\bar{u}_{01}+u_{0-1}\bar{u}_{00})\Psi_{14}+(|u_{0-1}|^2+|u_{00}|^2)\Psi_{24}],\vspace{0.08in}\\
 \Psi_{34,t}(t,k)=&2\alpha k(\bar{u}_{01}\Psi_{14}+\beta \bar{u}_{00}\Psi_{24})
      -i\alpha(\bar{u}_{11}\Psi_{14}+\beta \bar{u}_{10}\Psi_{24}) \vspace{0.08in}\\
    & +i\alpha[(|u_{01}|^2+|u_{00}|^2)\Psi_{34}+(\beta u_{0-1}\bar{u}_{00}+u_{00}\bar{u}_{01})\Psi_{44}] \vspace{0.08in}\\
 \Psi_{44,t}(t,k)=&2\alpha k(\bar{u}_{00}\Psi_{14}+\bar{u}_{0-1}\Psi_{24})
   -i\alpha(\bar{u}_{10}\Psi_{14}+\bar{u}_{1-1}\Psi_{24})\vspace{0.08in}\\
& +i\alpha[(\beta u_{00}\bar{u}_{0-1}+u_{01}\bar{u}_{00})\Psi_{34}+(|u_{0-1}|^2+|u_{00}|^2)\Psi_{44}],
\end{array}\right.
\ene

Suppose that $\Psi_{j1}$'s,\, $j=1,2,3,4$ are of the form
\bee\label{psi1}
\left(\begin{array}{c} \Psi_{11} \\ \Psi_{21} \\ \Psi_{31} \\ \Psi_{41} \end{array}\right)
=\left(a_{10}(t)+\frac{a_{11}(t)}{k}+\frac{a_{12}(t)}{k^2}+\cdots\right)+\left(b_{10}(t)+\frac{b_{11}(t)}{k}+\frac{b_{12}(t)}{k^2}+\cdots\right)e^{4ik^2t},
\ene
where the $4\times 1$ column vector functions $a_{1j}(t),b_{1j}(t)\, (j=0,1,...,)$ are independent of $k$.

By substituting Eq.~(\ref{psi1}) into Eq.(\ref{psit11}) and using the initial conditions
$a_{10}(0)+b_{10}(0)=(1, 0,0,0)^T, \, a_{11}(0)+b_{11}(0)=(0, 0, 0, 0)^T,$ we have
\bee\label{psi11}
\left(\begin{array}{c} \Psi_{11} \\ \Psi_{21} \\ \Psi_{31} \\ \Psi_{41} \end{array}\right)
=\left(\begin{array}{c} 1 \\ 0 \\ 0 \\ 0 \end{array}\right)
+\frac{1}{k}\left(\begin{array}{c} \Psi_{11}^{(1)} \vspace{0.05in}\\ \Psi_{21}^{(1)} \vspace{0.05in}\\ \Psi_{31}^{(1)} \vspace{0.05in}\\ \Psi_{41}^{(1)} \end{array}\right)
+\frac{1}{k^2}\left(\begin{array}{c} \Psi_{11}^{(2)} \vspace{0.05in}\\ \Psi_{21}^{(2)} \vspace{0.05in}\\ \Psi_{31}^{(2)} \vspace{0.05in}\\ \Psi_{41}^{(2)} \end{array}\right)
+O\left(\frac{1}{k^3}\right)
+\left[\frac{1}{k}\left(\begin{array}{c} 0 \\ 0 \vspace{0.05in}\\ -\frac{i\alpha}{2}\bar{u}_{01}(0) \vspace{0.05in}\\ -\frac{i\alpha}{2}\bar{u}_{00}(0) \end{array}\right)+O\left(\frac{1}{k^2}\right)  \right]e^{4ik^2t},
\ene

Similarly, it follows from Eqs.~(\ref{psit12})-(\ref{psit14}) that we have the asymptotic formulae for $\Psi_{ij},\, i=1,2,3,4; j=2,3,4$ in the form
\bee
\left(\begin{array}{c} \Psi_{12} \\ \Psi_{22} \\ \Psi_{32} \\ \Psi_{42} \end{array}\right)
=\left(\begin{array}{c} 0 \\ 1 \\ 0 \\ 0 \end{array}\right)
+\frac{1}{k}\left(\begin{array}{c} \Psi_{12}^{(1)} \vspace{0.05in}\\ \Psi_{22}^{(1)} \vspace{0.05in}\\ \Psi_{32}^{(1)} \vspace{0.05in}\\ \Psi_{42}^{(1)} \end{array}\right)
+\frac{1}{k^2}\left(\begin{array}{c} \Psi_{12}^{(2)} \vspace{0.05in}\\ \Psi_{22}^{(2)} \vspace{0.05in}\\ \Psi_{32}^{(2)} \vspace{0.05in}\\ \Psi_{42}^{(2)} \end{array}\right)
+O\left(\frac{1}{k^3}\right)
+\left[\frac{1}{k}\left(\begin{array}{c} 0 \\ 0 \vspace{0.05in}\\ -\frac{i\alpha\beta}{2}\bar{u}_{00}(0) \vspace{0.05in}\\ -\frac{i\alpha}{2}\bar{u}_{0-1}(0) \end{array}\right)+O\left(\frac{1}{k^2}\right)\right]e^{4ik^2t},
\label{psi12}
\ene
\bee
\left(\begin{array}{c} \Psi_{13} \\ \Psi_{23} \\ \Psi_{33} \\ \Psi_{43} \end{array}\right)
=\left(\begin{array}{c} 0 \\ 0 \\ 1 \\ 0 \end{array}\right)
+\frac{1}{k}\left(\begin{array}{c} \Psi_{13}^{(1)} \vspace{0.05in}\\ \Psi_{23}^{(1)} \vspace{0.05in}\\ \Psi_{33}^{(1)} \vspace{0.05in}\\ \Psi_{43}^{(1)} \end{array}\right)
+\frac{1}{k^2}\left(\begin{array}{c} \Psi_{13}^{(2)} \vspace{0.05in}\\ \Psi_{23}^{(2)} \vspace{0.05in}\\ \Psi_{33}^{(2)} \vspace{0.05in}\\ \Psi_{43}^{(2)} \end{array}\right)
+O\left(\frac{1}{k^3}\right)
+\left[\frac{1}{k}\left(\begin{array}{c} \frac{i}{2}u_{01}(0) \vspace{0.05in}\\  \frac{i\beta}{2}u_{00}(0) \vspace{0.05in}\\ 0 \vspace{0.05in}\\ 0\end{array}\right)+O\left(\frac{1}{k^2}\right)  \right]e^{-4ik^2t},
\label{psi13}
\ene
and
\bee
\left(\begin{array}{c} \Psi_{14} \\ \Psi_{24} \\ \Psi_{34} \\ \Psi_{44} \end{array}\right)
=\left(\begin{array}{c} 0 \\ 0 \\ 0 \\ 1 \end{array}\right)
+\frac{1}{k}\left(\begin{array}{c} \Psi_{14}^{(1)} \vspace{0.05in}\\ \Psi_{24}^{(1)} \vspace{0.05in}\\ \Psi_{34}^{(1)} \vspace{0.05in}\\ \Psi_{44}^{(1)} \end{array}\right)
+\frac{1}{k^2}\left(\begin{array}{c} \Psi_{14}^{(2)} \vspace{0.05in}\\ \Psi_{24}^{(2)} \vspace{0.05in}\\ \Psi_{34}^{(2)} \vspace{0.05in}\\ \Psi_{44}^{(2)} \end{array}\right)
+O\left(\frac{1}{k^3}\right)
+\left[\frac{1}{k}\left(\begin{array}{c} \frac{i}{2}u_{00}(0) \vspace{0.05in}\\  \frac{i}{2}u_{0-1}(0) \vspace{0.05in}\\ 0 \vspace{0.05in}\\ 0\end{array}\right)+O\left(\frac{1}{k^2}\right) \right]e^{-4ik^2t},
\label{psi14}
\ene

The substitution of Eqs.~(\ref{s13}) and (\ref{psi11})-(\ref{psi14}) into Eq.~(\ref{ca}) yields Eq.~(\ref{c13}). Similarly, we can also get Eqs.~(\ref{c14}) and (\ref{c24}).     $\square$

\subsection*{(c) \, The map between Dirichlet and Neumann problems}

\quad In the following we mainly show that the spectral functions $S(k)$ and $S_L(k)$ can be expressed in terms of the prescribed Dirichlet and Neumann boundary data and the initial data using the solution of a system of integral equations.

Define the new notations as
\bee\no
 F_{\pm} (t,k)=F(t,k)\pm F(t, -k), \quad \Sigma_{\pm}(k)=e^{2ikL}\pm e^{-2ikL}.
\ene
The sign $\partial D_j,\ j=1,...,4$ stands for the boundary of the $j$th quadrant $D_j$, oriented so that $D_j$ lies to the left of $\partial D_j$.
$\partial D_3^0$ denotes the boundary contour which has not contain the zeros of $\Sigma_-(k)$ and $\partial D_3^0=-\partial D_1^0$.

\vspace{0.1in}
\noindent{\bf Theorem 4.2.} {\it Let  the initial data of Eq.~(\ref{pnls}) $q_j(x,t=0)=q_{0j}(x),\, j=1,0,-1$ be the functions of Schwartz class
on the domain $x\in [0, \infty)$ and  $0<t<T<\infty$. For the Dirichlet problem, the boundary data $u_{0j}(t),\, (j=1,0,-1)$ on the interval $t\in [0, T)$ are sufficiently smooth and compatible with the initial data $q_{0j}(x),\, (j=1,0,-1)$ at the point $(x_2, t_2)=(0, 0)$, i.e., $u_{0j}(0)=q_{0j}(0),\, j=1,0,-1$. Similarly, for the Neumann problem, the boundary data $u_{1j}(t),\, j=1,0,-1$ on the interval $t\in [0, T)$ are sufficiently smooth and compatible with the initial data $q_{0j}(x),\, j=1,0,-1$ at the origin $(x_2, t_2)=(0, 0)$.}
{\it For simplicity, let $n_{33,44}(s)(k)$ have no zeros in the domain $D_1$. Then the matrix-valued spectral function $S(k)$ is
defined by
\bee \label{skm}
\begin{array}{l}S(k)=
\left(\!\!\!\begin{array}{llll}
 m_{11}(\Psi(T,k))  & \!\!  -m_{21}(\Psi(T,k)) & \!\!   m_{31}(\Psi(T,k))e^{4ik^2T} & \!\!   -m_{41}(\Psi(T,k))e^{4ik^2T} \vspace{0.05in}\\
 -m_{12}(\Psi(T,k)) & \!\!  m_{22}(\Psi(T,k)) & \!\!   -m_{32}(\Psi(T,k))e^{4ik^2T} & \!\!   m_{42}(\Psi(T,k))e^{4ik^2T} \vspace{0.05in}\\
 m_{13}(\Psi(T,k))e^{-4ik^2T}  & \!\!  -m_{23}(\Psi(T,k))e^{-4ik^2T} & \!\!   m_{33}(\Psi(T,k)) & \!\!   -m_{43}(\Psi(T,k)) \vspace{0.05in}\\
 -m_{14}(\Psi(T,k))e^{-4ik^2T} & \!\!  m_{24}(\Psi(T,k))e^{-4ik^2T} &  \!\!  -m_{34}(\Psi(T,k)) & \!\!   m_{44}(\Psi(T,k))
\end{array}\!\!\!\right),
\end{array}
\ene

and the complex-valued functions $\{\Psi_{ij}(t,k)\}_{i,j=1}^4$ have the following system of integral equations
\bee\label{psit10}
\left\{ \begin{array}{rl}
 \Psi_{11}(t,k)=&\!\!\! 1+\displaystyle\int_0^t \left\{-i\alpha[(|u_{01}|^2+|u_{00}|^2)\Psi_{11}+(\beta u_{01}\bar{u}_{02}+u_{00}\bar{u}_{0-1})\Psi_{21}]\right. \vspace{0.08in}\\
 &\displaystyle \qquad+\left. (2ku_{01}+iu_{11})\Psi_{31}+(2ku_{00}+iu_{10})\Psi_{41} \right\}(t',k)dt', \vspace{0.08in}\\
\Psi_{21}(t,k)=&\!\!\! \displaystyle\int_0^t
   \left\{-i\alpha[(\beta u_{00}\bar{u}_{01}+u_{0-1}\bar{u}_{00})\Psi_{11}+(|u_{0-1}|^2+|u_{00}|^2)\Psi_{21}]\right. \vspace{0.08in}\\
    &\qquad+\left. (2k(\beta u_{00}+i\beta u_{10})\Psi_{31}+ (2ku_{0-1}+iu_{1-1})\Psi_{41}\right\}(t',k)dt', \vspace{0.08in}\\
 \Psi_{31}(t,k)=&\!\!\!\displaystyle \int_0^te^{4ik^2(t-t')}\left\{2\alpha k(\bar{u}_{01}\Psi_{11}+\beta \bar{u}_{00}\Psi_{21})
      -i\alpha(\bar{u}_{11}\Psi_{11}+\beta \bar{u}_{10}\Psi_{21})\right. \vspace{0.08in}\\
  & \qquad\left.+i\alpha[(|u_{01}|^2+|u_{00}|^2)\Psi_{31}+(\beta u_{0-1}\bar{u}_{00}+u_{00}\bar{u}_{01})\Psi_{41}]\right\}
  (t',k)dt', \vspace{0.08in}\\
 \Psi_{41}(t,k)=&\!\!\! \displaystyle\int_0^te^{4ik^2(t-t')}\left[
  2\alpha k(\bar{u}_{00}\Psi_{11}+\bar{u}_{0-1}\Psi_{21})
   -i\alpha(\bar{u}_{10}\Psi_{11}+\bar{u}_{1-1}\Psi_{21})\right.\vspace{0.08in}\\
  & \qquad \displaystyle\left.+i\alpha[(\beta u_{00}\bar{u}_{0-1}+u_{01}\bar{u}_{00})\Psi_{31}+(|u_{0-1}|^2+|u_{00}|^2)\Psi_{41}]\right](t',k)dt',
\end{array}\right.
\ene

\bee\label{psit20}
\left\{ \begin{array}{rl}
 \Psi_{12}(t,k)=&\!\!\! \displaystyle\int_0^t
 \left\{-i\alpha[(|u_{01}|^2+|u_{00}|^2)\Psi_{12}+(\beta u_{01}\bar{u}_{02}+u_{00}\bar{u}_{0-1})\Psi_{22}]\right. \vspace{0.08in}\\
 & \displaystyle\left.+2k(u_{01}\Psi_{32}+u_{00}\Psi_{42})+i(u_{11}\Psi_{32}+u_{10}\Psi_{42})\right\}(t',k)dt',  \vspace{0.08in}\\
\Psi_{22}(t,k)=&\!\!\! \displaystyle 1+\int_0^t\left\{-i\alpha[(\beta u_{00}\bar{u}_{01}+u_{0-1}\bar{u}_{00})\Psi_{12}+(|u_{0-1}|^2+|u_{00}|^2)\Psi_{22}]\right.\vspace{0.08in}\\
& \displaystyle\left. +2k(\beta u_{00}\Psi_{32}+u_{0-1}\Psi_{42})+i(\beta u_{10}\Psi_{32}+u_{1-1}\Psi_{42})\right\}(t',k)dt', \vspace{0.08in}\\
 \Psi_{32}(t,k)=&\!\!\! \displaystyle\int_0^t e^{4ik^2(t-t')}\left\{2\alpha k(\bar{u}_{01}\Psi_{12}+\beta \bar{u}_{00}\Psi_{22})
      -i\alpha(\bar{u}_{11}\Psi_{12}+\beta \bar{u}_{10}\Psi_{22})\right. \vspace{0.08in}\\
& \left. \displaystyle+i\alpha[(|u_{01}|^2+|u_{00}|^2)\Psi_{32}+(\beta u_{0-1}\bar{u}_{00}+u_{00}\bar{u}_{01})\Psi_{42}]\right\}(t',k)dt', \vspace{0.08in}\\
 \Psi_{42}(t,k)=&\!\!\! \displaystyle\int_0^t e^{4ik^2(t-t')}\left\{
 2\alpha k(\bar{u}_{00}\Psi_{12}+\bar{u}_{0-1}\Psi_{22})
   -i\alpha(\bar{u}_{10}\Psi_{12}+\bar{u}_{1-1}\Psi_{22})\right.\vspace{0.08in}\\
& \displaystyle\left.+i\alpha[(\beta u_{00}\bar{u}_{0-1}+u_{01}\bar{u}_{00})\Psi_{32}+(|u_{0-1}|^2+|u_{00}|^2)\Psi_{42}]\right\}(t',k)dt',
\end{array}\right.
\ene

\bee\label{psit30}
\left\{ \begin{array}{rl}
 \Psi_{13}(t,k)=&\!\!\! \displaystyle\int_0^t
 e^{-4ik^2(t-t')}\left\{-i\alpha[(|u_{01}|^2+|u_{00}|^2)\Psi_{13}+(\beta u_{01}\bar{u}_{02}+u_{00}\bar{u}_{0-1})\Psi_{23}]\right., \vspace{0.08in}\\
 &\displaystyle\left.+2k(u_{01}\Psi_{33}+u_{00}\Psi_{43})+i(u_{11}\Psi_{33}+u_{10}\Psi_{43})\right\}(t',k)dt', \vspace{0.08in}\\
\Psi_{23}(t,k)=&\!\!\! \displaystyle\int_0^t
 e^{-4ik^2(t-t')}\left\{-i\alpha[(\beta u_{00}\bar{u}_{01}+u_{0-1}\bar{u}_{00})\Psi_{13}+(|u_{0-1}|^2+|u_{00}|^2)\Psi_{23}]\right.\vspace{0.08in}\\
&\displaystyle\left. +2k(\beta u_{00}\Psi_{33}+u_{0-1}\Psi_{43})+i(\beta u_{10}\Psi_{33}+u_{1-1}\Psi_{43})\right\}(t',k)dt', \vspace{0.08in}\\
 \Psi_{33}(t,k)=&\!\!\! 1+\displaystyle\int_0^t
 \left\{2\alpha k(\bar{u}_{01}\Psi_{13}+\beta \bar{u}_{00}\Psi_{23})
      -i\alpha(\bar{u}_{11}\Psi_{13}+\beta \bar{u}_{10}\Psi_{23})\right. \vspace{0.08in}\\
& \displaystyle\left.+i\alpha[(|u_{01}|^2+|u_{00}|^2)\Psi_{33}+(\beta u_{0-1}\bar{u}_{00}+u_{00}\bar{u}_{01})\Psi_{43}]\right\}(t',k)dt', \vspace{0.08in}\\
 \Psi_{43}(t,k)=&\!\!\! \displaystyle\int_0^t\left\{2\alpha k(\bar{u}_{00}\Psi_{13}+\bar{u}_{0-1}\Psi_{23})
   -i\alpha(\bar{u}_{10}\Psi_{13}+\bar{u}_{1-1}\Psi_{23})\right. \vspace{0.08in}\\
 & \displaystyle\left.+i\alpha[(\beta u_{00}\bar{u}_{0-1}+u_{01}\bar{u}_{00})\Psi_{33}+(|u_{0-1}|^2+|u_{00}|^2)\Psi_{43}]\right\}(t',k)dt',
\end{array}\right.
\ene
and
\bee\label{psit40}
\left\{ \begin{array}{rl}
 \Psi_{14}(t,k)=&\!\!\! \displaystyle\int_0^t
 e^{-4ik^2(t-t')}\left\{-i\alpha[(|u_{01}|^2+|u_{00}|^2)\Psi_{14}+(\beta u_{01}\bar{u}_{02}+u_{00}\bar{u}_{0-1})\Psi_{24}]\right. \vspace{0.08in}\\
 &\displaystyle \left.+2k(u_{01}\Psi_{34}+u_{00}\Psi_{44})+i(u_{11}\Psi_{34}+u_{10}\Psi_{44})\right\}(t',k)dt',  \vspace{0.08in}\\
   \Psi_{24}(t,k)=&\!\!\! \displaystyle\int_0^te^{-4ik^2(t-t')}\left\{-i\alpha[(\beta u_{00}\bar{u}_{01}+u_{0-1}\bar{u}_{00})\Psi_{14}+(|u_{0-1}|^2+|u_{00}|^2)\Psi_{24}]\right.\vspace{0.08in}\\
 &\displaystyle\left.+ 2k(\beta u_{00}\Psi_{34}+u_{0-1}\Psi_{44})+i(\beta u_{10}\Psi_{34}+u_{1-1}\Psi_{44})\right\}(t',k)dt', \vspace{0.08in}\\
 \Psi_{34}(t,k)=&\!\!\! \displaystyle\int_0^t
 \left[2\alpha k(\bar{u}_{01}\Psi_{14}+\beta \bar{u}_{00}\Psi_{24})
      -i\alpha(\bar{u}_{11}\Psi_{14}+\beta \bar{u}_{10}\Psi_{24})\right. \vspace{0.08in}\\
&\displaystyle \left.+i\alpha[(|u_{01}|^2+|u_{00}|^2)\Psi_{34}+(\beta u_{0-1}\bar{u}_{00}+u_{00}\bar{u}_{01})\Psi_{44}]\right\}(t',k)dt', \vspace{0.08in}\\
 \Psi_{44}(t,k)=&\!\!\! 1+\displaystyle\int_0^t\left\{
 2\alpha k(\bar{u}_{00}\Psi_{14}+\bar{u}_{0-1}\Psi_{24})
   -i\alpha(\bar{u}_{10}\Psi_{14}+\bar{u}_{1-1}\Psi_{24})\right.\vspace{0.08in}\\
&\displaystyle\left. +i\alpha[(\beta u_{00}\bar{u}_{0-1}+u_{01}\bar{u}_{00})\Psi_{34}+(|u_{0-1}|^2+|u_{00}|^2)\Psi_{44}]\right\}(t',k)dt',
\end{array}\right.
\ene }

(i) For the known Dirichlet problem, the unknown Neumann boundary conditions $u_{1j}(t),\, j=1, 0,-1,\, 0<t<T$ can be found by
\bes\bee\label{u11}
&\begin{array}{rl}
 u_{11}(t)=&\!\!\!\!\d\int_{\partial D_3}\!\!
\frac{2}{i\pi}\left[k\Psi_{13-}(t,-k)-iu_{01}(t)+u_{01}(t)\Psi_{33-}(t,k)+u_{00}(t)\Psi_{43-}(t,k)\right]dk \vspace{0.08in}\\
&\!\!\!\!\d +\frac{4i}{\pi}\int_{\partial D_3}
k\left\{[\Psi_{11}(t,k)s_{13}+\Psi_{12}(t,k)s_{23}]e^{-4ik^2t}\right.+\Psi_{13}(t,k)(s_{33}-1)+\Psi_{14}(t,k)s_{43}\Big\}dk, \qquad
\end{array} \vspace{0.08in}\\
\label{u10}
&\begin{array}{rl}
 u_{10}(t)=&\!\!\!\!\d\int_{\partial D_3}\!\!
\frac{2}{i\pi}\left[k\Psi_{14-}(t,-k)-iu_{00}(t)+u_{01}(t)\Psi_{34-}(t,k)+\beta u_{00}(t)\Psi_{44-}(t,k)\right]dk \vspace{0.08in}\\
&\!\!\!\!\d +\frac{4i}{\pi}\int_{\partial D_3}
k\left\{[\Psi_{11}(t,k)s_{14}+\Psi_{12}(t,k)s_{24}]e^{-4ik^2t}\right. +\Psi_{13}(t,k)s_{34}+\Psi_{14}(t,k)(s_{44}-1)\Big\}dk,  \qquad
\end{array} \vspace{0.08in}\\
\label{u1-1}
&\begin{array}{rl}
 u_{1-1}(t)=&\!\!\!\!\d\int_{\partial D_3}\!\!
\frac{2}{i\pi}\left[k\Psi_{24-}(t,-k)-iu_{0-1}(t)+u_{00}(t)\Psi_{34-}(t,k)+\beta u_{0-1}(t)\Psi_{44-}(t,k)\right]dk \vspace{0.08in}\\
&\!\!\!\!\d +\frac{4i}{\pi}\int_{\partial D_3}
k\left\{[\Psi_{21}(t,k)s_{14}+\Psi_{22}(t,k)s_{24}]e^{-4ik^2t}\right.+\Psi_{23}(t,k)s_{34}+\Psi_{24}(t,k)(s_{44}-1)\Big\}dk, \qquad
\end{array}
\ene\ees

(ii) For the known Neumann  problem, the unknown Dirichlet boundary conditions $u_{0j}(t),\, j=1, 0,-1,\, 0<t<T$ can be found by
\bes\bee \label{u01}
&\begin{array}{rl}
u_{01}(t)=&\!\! \d \frac{1}{\pi}\int_{\partial D_3^0}\Psi_{13+}(t, -k)dk -\frac{2}{\pi}\int_{\partial D_3}
\left\{[\Psi_{11}(t,k)s_{13}+\Psi_{12}(t,k)s_{23}]e^{-4ik^2t}\right. \vspace{0.08in}\\
&\!\!\! \d  +\Psi_{13}(t,k)(s_{33}-1)+\Psi_{14}(t,k)s_{43}\Big\}dk,
\end{array}\vspace{0.08in}\\
\label{u00}
&\begin{array}{rl}
 u_{00}(t)=&\!\! \d\frac{1}{\pi}\int_{\partial D_3^0}\Psi_{14+}(t, -k)dk -\frac{2}{\pi}\int_{\partial D_3}
\left\{[\Psi_{11}(t,k)s_{14}+\Psi_{12}(t,k)s_{24}]e^{-4ik^2t}\right. \vspace{0.08in}\\
&\!\!\! \d +\Psi_{13}(t,k)s_{34}+\Psi_{14}(t,k)(s_{44}-1)\Big\}dk,  \qquad
\end{array} \vspace{0.08in}\\
\label{u0-1}
&\begin{array}{rl}
 u_{1-1}(t)=&\!\!\d \frac{1}{\pi}\int_{\partial D_3^0}\Psi_{24+}(t, -k)dk -\frac{2}{\pi}\int_{\partial D_3}
\left\{[\Psi_{21}(t,k)s_{14}+\Psi_{22}(t,k)s_{24}]e^{-4ik^2t}\right. \vspace{0.08in}\\
&\!\! \d +\Psi_{23}(t,k)s_{34}+\Psi_{24}(t,k)(s_{44}-1)\Big\}dk,  \qquad
\end{array}
\ene\ees
where $s_{ij}=s_{ij}(k),\, i,j=1,2,3,4$.

\vspace{0.1in}

\noindent {\bf Proof.} We can show Eq.~(\ref{skm}) by means of Eq.~(\ref{sss}), that is,
 \bee
   S(k)=e^{-2ik^2T\hat{\sigma}_4}\mu_2^{-1}(0,T,k)=e^{-2ik^2T\hat{\sigma}_4}\Big(\mu_2^A(0,T,k)\Big)^T
   =e^{-2ik^2T\hat{\sigma}_4}\Big(\Psi^A(T,k)\Big)^T,
    \ene
Moreover, Eqs.~(\ref{psit10})-(\ref{psit40}) for $\Psi_{ij}(t,k),\, i,j=1,2,3,4$ can be obtained by using the Volteral integral equations of $\mu_2(0,t,k)$.

\vspace{0.1in}
(i) In the following we show Eqs.~(\ref{u11})-(\ref{u1-1}). Applying the Cauchy's theorem to Eq.~(\ref{mu2asy}), we have
\bee\label{psit}
\begin{array}{l}
-\dfrac{i\pi }{2}\Psi_{33}^{(1)}(t)=\displaystyle\int_{\partial D_2}[\Psi_{33}(t,k)-1]dk=\int_{\partial D_4}[\Psi_{33}(t,k)-1]dk,
 \vspace{0.08in}\\
-\dfrac{i\pi }{2}\Psi_{43}^{(1)}(t)=\displaystyle\int_{\partial D_2}\Psi_{43}(t,k)dk=\int_{\partial D_4}\Psi_{43}(t,k)dk, \vspace{0.08in}\\
-\dfrac{i\pi }{2}\Psi_{13}^{(2)}(t)=\displaystyle\int_{\partial D_2}\left[k\Psi_{13}(t,k)+\frac{i}{2}u_{01}(t)\right]dk=-\int_{\partial D_4}\left[k\Psi_{13}(t,k)+\frac{i}{2}u_{01}(t)\right]dk,
\end{array}\ene

From Eq.~(\ref{psit}), we further find
\bes\bee
&\label{psi33}
\begin{array}{rl}
i\pi \Psi_{33}^{(1)}(t)=& \d -\left(\int_{\partial D_2}+\int_{\partial D_4}\right)[\Psi_{33}(t,k)-1]dk = \left(\int_{\partial D_1}+\int_{\partial D_3}\right)[\Psi_{33}(t,k)-1]dk \vspace{0.08in}\\
=&  \d\int_{\partial D_3}[\Psi_{33}(t,k)-1]dk-\int_{\partial D_3}[\Psi_{33}(t,-k)-1]dk = \d \int_{\partial D_3}\Psi_{33-}(t,k)dk,
\end{array} \\
&\label{psi43}
i\pi \Psi_{43}^{(1)}(t)=\d \int_{\partial D_3}\Psi_{43-}(t,k)dk,
\ene\ees
\bee\label{psi13}
\begin{array}{rl}
i\pi \Psi_{13}^{(2)}(t)=&\!\!\! \d\left(\int_{\partial D_1}-\int_{\partial D_3}\right)\left[k\Psi_{13}(t,k)+\frac{i}{2}u_{01}(t)\right]dk+C_1(t)=\int_{\partial D_3}\left[k\Psi_{13-}(t,-k)-i u_{01}(t)\right]dk+C_1(t),
\end{array}
\ene
where we have introduced the function $C_1(t)$ in the form
\bee\no
C_1(t)=\d 2\int_{\partial D_3}\left[k\Psi_{13}(t,k)+\frac{i}{2}u_{01}(t)\right]dk,
\ene

We use the global relation (\ref{ca}), the Cauchy's theorem and asymptotic (\ref{c13}) to further reduce $C_1(t)$ to be
\bee
\label{c1}
\begin{array}{rl}
C_1(t)=&\!\!\d 2\int_{\partial D_3}
\left[kc_{13}(t,k)+\frac{i}{2}u_{01}(t)\right]dk-\d 2\int_{\partial D_3}k\left[c_{13}(t,k)-\Psi_{13}(t,k)\right]dk,  \vspace{0.08in}\\
=&\!\!\d-i\pi\Psi_{13}^{(2)}\!-\!2\int_{\partial D_3}\!k\left\{[\Psi_{11}(t,k)s_{13}(k)\!+\!\Psi_{12}(t,k)s_{23}(k)]e^{-4ik^2t}\right.   \!+\!\Psi_{13}(t,k)(s_{33}(k)\!-\!1)\!+\!\Psi_{14}(t,k)s_{43}(k)\Big\}dk,
\end{array}
\ene

It follows from Eqs.~(\ref{psi13}) and (\ref{c1}) that we have
\bee\label{psi13-}
\begin{array}{rl}
2i\pi \Psi_{13}^{(2)}(t)=&\!\!\!\!\! \d\int_{\partial D_3}\left[k\Psi_{13-}(t,-k)-i u_{01}(t)\right]dk  -2\int_{\partial D_3}k\left\{[\Psi_{11}(t,k)s_{13}(k)+\Psi_{12}(t,k)s_{23}(k)]e^{-4ik^2t}\right.  \vspace{0.08in}\\
& \d  +\Psi_{13}(t,k)(s_{33}(k)-1)+\Psi_{14}(t,k)s_{43}(k)\Big\}dk,
\end{array}\ene
Thus substituting Eqs.~(\ref{psi33}), (\ref{psi43}) and (\ref{psi13-}) into the fourth one of system (\ref{ud}), we can get Eq.~(\ref{u11}).
Similarly, we can also show that Eqs.~(\ref{u10}) and (\ref{u1-1}) hold.

\vspace{0.1in}
(ii) We now derive the Dirichlet boundary value conditions (\ref{u01})-(\ref{u0-1}) at $x=0$ from the given Neumann boundary value problems. It follows from the first one of Eq.~(\ref{ud}) that $u_{01}(t)$ can be expressed by means of $\Psi_{13}^{(1)}$.
Applying the Cauchy's theorem to Eq.~(\ref{mu2asy}) yields
\bee\label{psi13+}
\begin{array}{rl}
i\pi \Psi_{13}^{(1)}(t)=&\!\!\! \d\left(\int_{\partial D_1}+\int_{\partial D_3}\right)\Psi_{13}(t,k)dk= \d\left(\int_{\partial D_1}-\int_{\partial D_3}\right)\Psi_{13}(t,k)dk+C_2(t) \vspace{0.08in}\\
=&\!\!\! \d\int_{\partial D_1}\Psi_{13+}(t,k)dk+C_2(t)= \d\int_{\partial D_3}\Psi_{13+}(t,-k)dk+C_2(t),
\end{array}
\ene
where we have introduced the function $C_2(t)$ in the form
\bee\no
C_2(t)=\d 2\int_{\partial D_3}\Psi_{13}(t,k)dk,
\ene

We use the global relation (\ref{ca}),  the Cauchy's theorem and asymptotics (\ref{c13}) to further reduce $C_2(t)$ to be
\bee\label{c2}
\begin{array}{rl}
C_2(t)=&\!\!\d 2\int_{\partial D_3}c_{13}(t,k)dk-\d 2\int_{\partial D_3}\left[c_{13}(t,k)-\Psi_{13}(t,k)\right]dk, \vspace{0.08in}\\
=&\!\!\d-i\pi\Psi_{13}^{(1)}\!-\!2\int_{\partial D_3}\!\left\{[\Psi_{11}(t,k)s_{13}(k)\!+\!\Psi_{12}(t,k)s_{23}(k)]e^{-4ik^2t}\right.  \!+\!\Psi_{13}(t,k)(s_{33}(k)-1)\!+\!\Psi_{14}(t,k)s_{43}(k)\Big\}dk,
\end{array}
\ene

Eqs.~(\ref{psi13+}) and (\ref{c2}) imply that
\bee \label{psi13g2g}
\begin{array}{rl}
2i\pi \Psi_{13}^{(1)}(t)=&\!\! \d\int_{\partial D_3}\Psi_{13+}(t,-k)dk-2\int_{\partial D_3}\left\{[\Psi_{11}(t,k)s_{13}(k)+\Psi_{12}(t,k)s_{23}(k)]e^{-4ik^2t}\right. \vspace{0.08in}\\
& \d  +\Psi_{13}(t,k)(s_{33}(k)-1)+\Psi_{14}(t,k)s_{43}(k)\Big\}dk,
\end{array}
\ene
Thus, the substitution of Eq.~(\ref{psi13g2g}) into the first one of Eq.~(\ref{ud}) yields Eq.~(\ref{u01}). Similarly, in terms of the global relation (\ref{cb}) and (\ref{cc}), we can also show Eqs.~(\ref{u00}) and (\ref{u0-1}) by using the second and third ones of Eq.~(\ref{ud}) and $\Psi_{14}^{(1)}(t)$ and $\Psi_{24}^{(1)}(t)$.  $\square$

\subsection*{(d) \, Effective characterizations }

\quad Substituting the perturbated expressions of the eigenfunction and initial-boundary data
\bee\label{solue}
\begin{array}{l}
\Psi_{ij}(t,k)=\Psi_{ij}^{[0]}+\epsilon\Psi_{ij}^{[1]}+\epsilon^2\Psi_{ij}^{[2]}+\cdots, \quad  i,j=1,2,3,4, \vspace{0.08in}\\
u_{0j}(t)=\epsilon u_{0j}^{[1]}(t)+\epsilon^2 u_{0j}^{[2]}(t)+\cdots, \quad  j=1,0,-1, \vspace{0.08in}\\
u_{1j}(t)=\epsilon u_{1j}^{[1]}(t)+\epsilon^2 u_{1j}^{[2]}(t)+\cdots, \quad  j=1,0,-1,
\end{array}
\ene
where $\epsilon>0$ is a small parameter,  into Eqs.~(\ref{psit11})-(\ref{psit14}), we have these terms of $O(1)$ and
$O(\epsilon)$ as
\bee
O(1): \left\{\begin{array}{l}
 \Psi_{jj}^{[0]}(t,k)=1,\quad j=1,2,3,4, \vspace{0.1in}\\
 \Psi_{ij}^{[0]}(t,k)=0,\quad i,j=1,2,3,4,\, i\not=j,
\end{array}\right. \qquad\qquad\qquad\qquad\qquad\qquad\quad
\ene
\bee \label{epsilon}
O(\epsilon): \left\{\begin{array}{l}
 \Psi_{11}^{[1]}(t,k)=\Psi_{12}^{[1]}=\Psi_{21}^{[1]}=\Psi_{22}^{[1]}=
 \Psi_{33}^{[1]}(t,k)=\Psi_{34}^{[1]}=\Psi_{43}^{[1]}=\Psi_{44}^{[1]}=0, \vspace{0.1in}\\
 \Psi_{13}^{[1]}(t,k)=\d\int_0^te^{-4ik^2(t-t')}\left(2ku_{01}^{[1]}+iu_{11}^{[1]}\right)(t')dt',
 \Psi_{14}^{[1]}(t,k)=\d\int_0^te^{-4ik^2(t-t')}\left(2ku_{00}^{[1]}+iu_{10}^{[1]}\right)(t')dt',\vspace{0.1in}\\
 \Psi_{23}^{[1]}(t,k)=\d\beta\int_0^te^{-4ik^2(t-t')}\left(2ku_{00}^{[1]}+iu_{10}^{[1]}\right)(t')dt',
 \Psi_{24}^{[1]}(t,k)=\d\int_0^te^{-4ik^2(t-t')}\left(2ku_{0-1}^{[1]}+iu_{1-1}^{[1]}\right)(t')dt',\vspace{0.1in}\\
 \Psi_{31}^{[1]}(t,k)=\d\alpha\int_0^te^{4ik^2(t-t')}\left(2k\bar{u}_{01}^{[1]}-i\bar{u}_{11}^{[1]}\right)(t')dt',
  \Psi_{32}^{[1]}(t,k)=\d\alpha\beta\int_0^te^{4ik^2(t-t')}\left(2k\bar{u}_{00}^{[1]}-i\bar{u}_{10}^{[1]}\right)(t')dt',\vspace{0.1in}\\
 \Psi_{41}^{[1]}(t,k)=\d\alpha\int_0^te^{4ik^2(t-t')}\left(2k\bar{u}_{00}^{[1]}-i\bar{u}_{10}^{[1]}\right)(t')dt',
 \Psi_{42}^{[1]}(t,k)=\d\alpha\int_0^te^{4ik^2(t-t')}\left(2k\bar{u}_{0-1}^{[1]}-i\bar{u}_{1-1}^{[1]}\right)(t')dt',
 \end{array}\right.
\ene

If we assume that $n_{33,44}(s)$ has no zeros, then we substitute the fourth one in Eq.~(\ref{solue}) into Eqs.~(\ref{u11})-(\ref{u1-1}) to find
\bee \label{u0111}\left\{\begin{array}{l}
 u_{11}^{[1]}(t)=\d\int_{\partial D_3}\left[\frac{2}{i\pi}\left(k\Psi_{13-}^{[1]}(t, -k)-iu_{01}^{[1]}\right)
  +\frac{4ik}{\pi} s_{13}^{[1]}\right]dk, \vspace{0.08in}\\
 u_{10}^{[1]}(t)=\d\int_{\partial D_3}\left[\frac{2}{i\pi}\left(k\Psi_{14-}^{[1]}(t, -k)-iu_{00}^{[1]}\right)
  +\frac{4ik}{\pi} s_{14}^{[1]}\right]dk, \vspace{0.08in}\\
 u_{1-1}^{[1]}(t)=\d\int_{\partial D_3}\left[\frac{2}{i\pi}\left(k\Psi_{24-}^{[1]}(t, -k)-iu_{0-1}^{[1]}\right)
  +\frac{4ik}{\pi} s_{24}^{[1]}\right]dk,
\end{array}\right.
\ene
where $s_{13}=\epsilon s_{13}^{[1]}(t)+\epsilon^2 s_{13}^{[2]}(t)+O(\epsilon^3),$ $s_{14}=\epsilon s_{14}^{[1]}(t)+\epsilon^2 s_{14}^{[2]}(t)+O(\epsilon^3),$ and  $s_{24}=\epsilon s_{24}^{[1]}(t)+\epsilon^2 s_{24}^{[2]}(t)+O(\epsilon^3)$.

It further follows from Eq.~(\ref{epsilon}) that we have
\bee\label{psi-}
 \left\{\begin{array}{l}
 \Psi_{13-}^{[1]}(t, -k)=\d -4k\int_0^te^{-4ik^2(t-t')}u_{01}^{[1]}(t')dt',\vspace{0.1in}\\
 \Psi_{14-}^{[1]}(t, -k)=\d -4k\int_0^te^{-4ik^2(t-t')}u_{00}^{[1]}(t')dt',\vspace{0.1in}\\
 \Psi_{24-}^{[1]}(t, -k)=\d -4k\int_0^te^{-4ik^2(t-t')}u_{0-1}^{[1]}(t')dt',
\end{array}\right.
\ene

Thus, the Dirichlet problem can now be solved perturbatively as follows: for $n_{33,44}(s)$ having no zeros
and given $u_{0j}^{[1]}, \, j=1,0,-1$, we can obtain $\{\Psi_{ij-}^{[1]},\, i=12; j=3,4$ from Eq.~(\ref{psi-})
and further find $u_{1j}^{[1]},\, j=1,0,-1$ from Eq.~(\ref{u0111}). Finally, we can have $\Psi_{ij}^{[1]}$ from Eq.~(\ref{epsilon}).
In fact, these arguments for $\Psi_{ij}$ can be extended to all orders such that we can determine all orders of $S(k)$.

In fact, the above recursive formulae can be continued indefinitely. We assume that they hold for all $0\leq j\leq n-1$, then for $n>0$,
the substitution of Eq.~(\ref{solue}) into Eqs.~(\ref{u11})-(\ref{u1-1}) yields the terms of $O(\epsilon^n)$ as
\bes\bee\label{u11e}
u_{11}^{[n]}(t)=\d\int_{\partial D_3}\left[\frac{2}{i\pi}\left(k\Psi_{13-}^{[n]}(t, -k)-iu_{01}^{[n]}\right)
  +\frac{4ik}{\pi} s_{13}^{[n]}\right]dk+ {\rm lower \,\, order \,\, terms}, \\
\label{u10e}
 u_{10}^{[n]}(t)=\d\int_{\partial D_3}\left[\frac{2}{i\pi}\left(k\Psi_{14-}^{[n]}(t, -k)-iu_{00}^{[n]}\right)
  +\frac{4ik}{\pi} s_{14}^{[n]}\right]dk+ {\rm lower \,\, order \,\, terms},\\
  \label{u1-1e}
 u_{1-1}^{[n]}(t)=\d\int_{\partial D_3}\left[\frac{2}{i\pi}\left(k\Psi_{24-}^{[n]}(t, -k)-iu_{0-1}^{[n]}\right)
  +\frac{4ik}{\pi} s_{24}^{[n]}\right]dk+ {\rm lower \,\, order \,\, terms},
\ene\ees
where `lower order terms' stands for the result involving known terms of lower order.

The terms of $O(\epsilon^n)$ in Eqs.~(\ref{psit10})-(\ref{psit40}) yield
\bee\label{psin}
 \left\{\begin{array}{l}
  \Psi_{13}^{[n]}(t, k)=\d\int_0^te^{-4ik^2(t-t')}\left(2ku_{01}^{[n]}+iu_{11}^{[n]}\right)(t')dt' + {\rm lower \,\, order \,\, terms},\vspace{0.1in}\\
 \Psi_{14}^{[n]}(t, k)=\d\int_0^te^{-4ik^2(t-t')}\left(2ku_{00}^{[n]}+iu_{10}^{[n]}\right)(t')dt' + {\rm lower \,\, order \,\, terms},\vspace{0.1in}\\
 \Psi_{24}^{[n]}(t, k)=\d\int_0^te^{-4ik^2(t-t')}\left(2ku_{0-1}^{[n]}+iu_{1-1}^{[n]}\right)(t')dt' + {\rm lower \,\, order \,\, terms}, \end{array}\right.
\ene
which leads to
\bee\label{psin-}
 \left\{\begin{array}{l}
 \Psi_{13-}^{[n]}(t, -k)=\d -4k\int_0^te^{-4ik^2(t-t')}u_{01}^{[n]}(t')dt'+ {\rm lower \,\, order \,\, terms},\vspace{0.1in}\\
 \Psi_{14-}^{[n]}(t, -k)=\d -4k\int_0^te^{-4ik^2(t-t')}u_{00}^{[n]}(t')dt'+ {\rm lower \,\, order \,\, terms},\vspace{0.1in}\\
 \Psi_{24-}^{[n]}(t, -k)=\d -4k\int_0^te^{-4ik^2(t-t')}u_{0-1}^{[n]}(t')dt'+ {\rm lower \,\, order \,\, terms},
\end{array}\right.
\ene
It follows from system (\ref{psin-}) that $\Psi_{13-}^{[n]}(t, -k),\, \Psi_{14-}^{[n]}(t, -k),$ and $\Psi_{24-}^{[n]}(t, -k)$ can be generated at each step from the known Dirichlet boundary data $u_{0j}^{[n]}(t),\, j=1,0,-1$ such that we know that the Neumann boundary data $u_{1j}^{[n]}(t),\, j=1,0,-1$ can be given by Eqs.~(\ref{u11e})-(\ref{u1-1e}) and then $\Psi_{13}^{[n]}(t, k),\, \Psi_{14}^{[n]}(t, k),$ and $\Psi_{24}^{[n]}(t,k)$ can be determined by Eq.~(\ref{psin}) and other $\Psi_{ij}^{[n]}(t,k)$ can also be found.

Similarly, it follows from Eqs.~(\ref{u01})-(\ref{u0-1}) that we have
\bee \label{u1101}
&\begin{array}{l}
u_{01}^{[1]}(t)=\d \frac{1}{\pi}\int_{\partial D_3}\left[\Psi_{13+}^{[1]}(t,-k)-2s_{13}^{[1]}\right]dk, \vspace{0.08in}\\
u_{00}^{[1]}(t)=\d \frac{1}{\pi}\int_{\partial D_3}\left[\Psi_{14+}^{[1]}(t,-k)-2s_{14}^{[1]}\right]dk, \vspace{0.08in}\\
u_{0-1}^{[1]}(t)=\d \frac{1}{\pi}\int_{\partial D_3}\left[\Psi_{24+}^{[1]}(t,-k)-2s_{24}^{[1]}\right]dk,
\end{array}
\ene

It further follows from Eq.~(\ref{epsilon}) that we have
\bee \label{psi+}
\left\{\begin{array}{l}
  \Psi_{13+}^{[1]}(t,-k)=\d 2i\int_0^te^{-4ik^2(t-t')}u_{11}^{[1]}(t')dt',\vspace{0.1in}\\
 \Psi_{14+}^{[1]}(t,-k)=\d  2i\int_0^te^{-4ik^2(t-t')}u_{10}^{[1]}(t')dt',\vspace{0.1in}\\
 \Psi_{24+}^{[1]}(t,-k)=\d  2i\int_0^te^{-4ik^2(t-t')}u_{1-1}^{[1]}(t')dt',
 \end{array} \right.
\ene
Thus, the Neumann problem can now be solved perturbatively as follows: for $n_{33,44}(s)$ having no zeros
and given $u_{1j}^{[1]}(t),\, j=1,0,-1$, we can obtain $\{\Psi_{13+}^{[1]},\, \Psi_{14+}^{[1]},\, \Psi_{24+}^{[1]}\}$
from Eq.~(\ref{psi+}) and further find $u_{0j}^{[1]},\, j=1,0,-1$ from Eq.~(\ref{u1101}). Finally, we can have $\Psi_{ij}^{[1]}$ from Eq.~(\ref{epsilon}). In fact, these arguments for $\Psi_{ij}$  can be extended to all orders such that we can determine all orders of $S(k)$.

Similarly, the substitution of Eq.~(\ref{solue}) into Eqs.~(\ref{u01})-(\ref{u0-1}) yields the terms of $O(\epsilon^n)$ as
\bes\bee \label{u01e}
& u_{01}^{[n]}(t)=\d \int_{\partial D_3}\left[\Psi_{13+}^{[n]}(t,-k)-2s_{13}^{[n]}\right]dk+ {\rm lower \,\, order \,\, terms},\\
& \label{u00e}
u_{00}^{[n]}(t)=\d \int_{\partial D_3}\left[\Psi_{14+}^{[n]}(t,-k)-2s_{14}^{[n]}\right]dk+ {\rm lower \,\, order \,\, terms},\\
& \label{u0-1e}
u_{0-1}^{[n]}(t)=\d \int_{\partial D_3}\left[\Psi_{24+}^{[n]}(t,-k)-2s_{24}^{[n]}\right]dk+ {\rm lower \,\, order \,\, terms},
\ene\ees

Eq.~(\ref{psin}) implies that
\bee \label{psin+}
\left\{\begin{array}{l}
  \Psi_{13+}^{[n]}(t,-k)=\d 2i\int_0^te^{-4ik^2(t-t')}u_{11}^{[n]}(t')dt'+ {\rm lower \,\, order \,\, terms},\vspace{0.1in}\\
 \Psi_{14+}^{[n]}(t,-k)=\d  2i\int_0^te^{-4ik^2(t-t')}u_{10}^{[n]}(t')dt'+ {\rm lower \,\, order \,\, terms},\vspace{0.1in}\\
 \Psi_{24+}^{[n]}(t,-k)=\d  2i\int_0^te^{-4ik^2(t-t')}u_{1-1}^{[n]}(t')dt'+ {\rm lower \,\, order \,\, terms}, \end{array} \right.
\ene
It follows from system (\ref{psin+}) that $\Psi_{13+}^{[n]},\, \Psi_{14+}^{[n]},\, \Psi_{24+}^{[n]}$  can be generated at each step from the known Neumann boundary data $u_{1j}^{[n]},\, j=1,0,-1$ such that we know that the Dirichlet boundary data $u_{0j}^{[n]},\, j=1,0,-1$ can then be given
by Eqs.~(\ref{u01e})-(\ref{u0-1e}).

\vspace{0.1in}
\noindent {\bf Remark 4.3}. We can also give the corresponding Gelfand-Levitan-Marchenko representations for the Dirichlet and Neumann boundary value problems and the IVB of spin-1 GP equation on the finite interval, which will be studied in another paper. The analogous analysis of the Fokas unified method can also be extended to study the IBV problems for other integrable nonlinear evolution PDEs with $4\times 4$ Lax pairs on the half-line or the finite interval.

\vspace{0.1in}
\noindent
{\bf Acknowledgment}
\vspace{0.1in}
 This work was partially supported by the NSFC under Grant No.11571346 and the Youth Innovation Promotion Association, CAS.

\end{document}